\documentclass[aps,reprint, prl, superscriptaddress, longbibliography,showpacs,nobibnotes,floatfix]{revtex4-1}
\usepackage{graphicx}
\usepackage{bm, amsmath,amssymb}
\usepackage{dcolumn}
\usepackage{natbib,xcolor}
\usepackage[version=4]{mhchem}
\usepackage{upgreek}
\usepackage{physics}
\usepackage{appendix}

\begin{document}
\title{Symmetry-based requirement for the measurement of electrical and thermal Hall conductivity under an in-plane magnetic field}

\author{Takashi Kurumaji}
\affiliation{Department of Advanced Materials Science, University of Tokyo, Kashiwa 277-8561, Japan}
\email{kurumaji@edu.k.u-tokyo.ac.jp}

\date{\today}
\begin{abstract}
The in-plane (thermal) Hall effect is an unconventional transverse response when the applied magnetic field is in the (heat) current plane.
In contrast to the normal Hall effect, the in-plane Hall effect requires the absence of certain crystal symmetries, and possibly manifests a non-trivial topology of quantum materials.
An accurate estimation of the intrinsic in-plane (thermal) Hall conductivity is crucial to identify the underlying mechanisms as in the case of the Kitaev spin-liquid candidate $\alpha$-\ce{RuCl3}.
Here, we give the symmetry conditions for the in-plane Hall effect and discuss the implications that may impede the experimental evaluation of the in-plane (thermal) Hall conductivity within the single-device measurement.
First, the lack of symmetry in crystals can create merohedral twin domains that cancel the total Hall signal.
Second, even in a twin-free crystal, the intrinsic response is potentially contaminated by the out-of-plane conduction in three-dimensional systems, which is systematically unavoidable in the in-plane Hall systems.
Third, even in a quasi-two-dimensional system, the conversion of (thermal) resistivity, $\hat{\rho}$ ($\hat{\lambda}$), to (thermal) conductivity, $\hat{\sigma}$ ($\hat{\kappa}$) requires protocols beyond the widely-used simplified formula $\sigma_{xy}=\rho_{yx}/(\rho_{xx}^2+\rho_{yx}^2)$ ($\kappa_{xy}=\lambda_{yx}/(\lambda_{xx}^2+\lambda_{yx}^2)$) due to the lack of in-plane-rotational symmetry.
In principle, two independent sample devices are necessary to accurately estimate the $\sigma_{xy}$ ($\kappa_{xy}$).
As a case study, we discuss the half-integer quantization of the in-plane thermal Hall effect in the spin-disordered state of $\alpha$-\ce{RuCl3}.
For an accurate measurement of the thermal Hall effect, it is necessary to avoid crystals with the merohedral twins contributing oppositely to $\kappa_{xy}$, while the out-of-plane transport may have a negligible effect.
To deal with the field-induced rotational-symmetry breaking, we propose two symmetry-based protocols, improved single-device and two-device methods.
The considerations in the manuscript are generally applicable to a broad class of materials and provide a useful starting point for understanding the unconventional aspects of the in-plane Hall effect.
\end{abstract}

\maketitle
\section{Introduction.}
The conventional Hall effect occurs in metals and semiconductors as a transverse electric field ($E_y$) in an electric current along the $x$ axis ($J_x$) when the magnetic field ($B$) is applied along the $z$ direction (Fig. 1(a)) \cite{hurd2012hall}.
This is due to the Lorentz force on the conduction electrons and the $E_y$ changes its sign when the $B_z$ is reversed ($E_y(-B_z)=-E_y(B_z)$).
Two other types of transverse responses are known, which are induced when the magnetic field is applied in the $xy$ plane.
One of them is the planar Hall effect, which occurs when $B$ is rotated by an angle $\theta$ from the $x$ axis to the $y$ axis (Fig. 1(b)) \cite{goldberg1954new,koch1955problem,ky1966plane}.
This is actually not a true field-odd Hall effect, but rather a field-even response resulting from the anisotropy of magnetoresistance \cite{jan1957galvamomagnetic}.
The other is the in-plane Hall effect, which is the main interest of this study.
The typical configuration is shown in Fig. 1(c), where the applied $B_x$ is parallel to $J_x$ and the induced $E_y$ is reversed by the inversion of $B$, and thus is the field-odd response \footnote{Equivalently, the in-plane Hall effect is defined as the Hall electric field parallel to the magnetic field ($B_x\parallel E_x$) when the current ($J_y$) is applied perpendicular to $B_x$ (see Fig. \ref{figFive}(c)).}.
This effect has been reported experimentally in cubic germanium as the $B^3$-order effect \cite{garcia1959variational,grabner1960longitudinal}, as well as in trigonal bismuth \cite{okada1956measurement,okada1957measurements}, and in monoclinic binary semimetals as the $B$-linear effect \cite{mollendorf1984first,bauhofer1988two}.

\begin{figure}[t]
	\includegraphics[width =  \columnwidth]{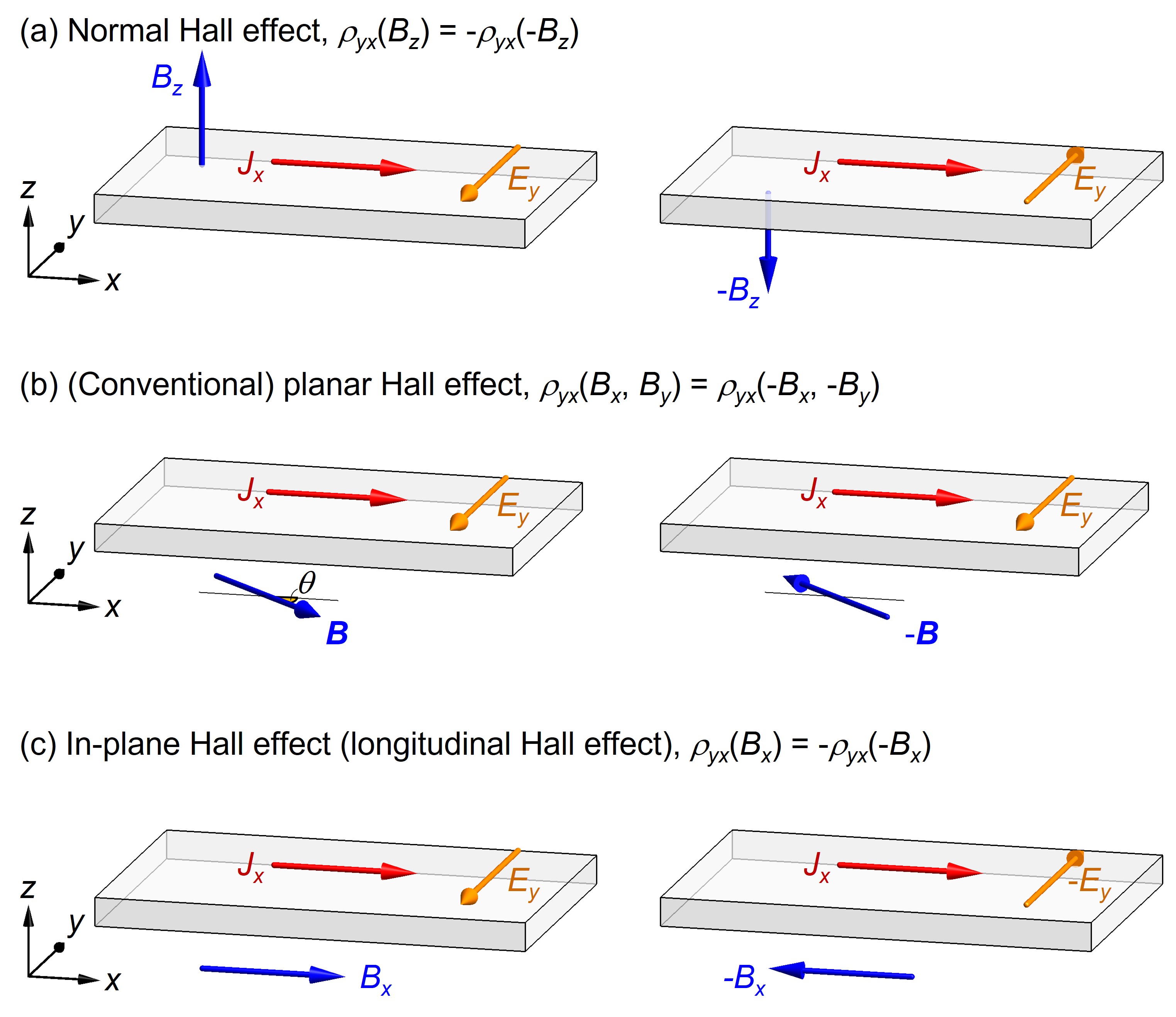}
	\caption{\label{figHall}
 Schematic configurations of (a) normal Hall effect, (b) conventional planar Hall effect, and (c) field-odd in-plane Hall effect (longitudinal Hall effect).
 Gray rectangle is a crystal.
 Blue, red, and orange arrows represent the magnetic field $B$, current $J_x$ applied along the $x$ axis, and the induced Hall electric field $E_y$ along the $y$ axis, respectively.
 $xyz$ is the Cartesian coordinate.
 Left and right panels represent the $E_y$ response to the $B$-field reversal.}
\end{figure}

The in-plane Hall effect, or sometimes referred to as the longitudinal Hall effect, was previously recognized as a consequence of the multi-band effect \cite{grabner1960longitudinal,bresler1972galvanomagnetic,akgoz1974galvanomagnetic,bauhofer1985longitudinal,bauhofer1988two,baltz1993theoretical,klar1994galvanomagnetic}.
Independently from these early studies, this effect has recently attracted increasing interest in the context of a topological signature of quantum materials.
Theories consider the quantization of the anomalous Hall effect by in-plane magnetization on magnetic topological insulators \cite{zhang2011quantized,liu2013plane,ren2016quantum,sheng2017monolayer,zhong2017plane,zhang2019plane,liu2018intrinsic,li2022chern}, the effect of the spin-orbit interaction \cite{mal1998hall,zyuzin2020plane}, and the field-induced quantization in two-dimensional electron gas \cite{zyuzin2020plane,battilomo2021anomalous,sun2022possible}.
The role of the Berry curvature in the electron bands suggests an intrinsic (dissipationless) nature of the in-plane Hall effect in topologically non-trivial semimetals \cite{zyuzin2020plane,cullen2021generating,tan2021unconventional,wang2022theory}.
Experimentally, the in-plane Hall conductivity is suggested to be a signature of unconventional topological transports as reported in the nonmagnetic semimetal \ce{ZrTe5} \cite{liang2018anomalous}, heterodimensional superlattice of \ce{VS2}-\ce{VS} \cite{zhou2022heterodimensional}, and magnetic half-Heusler DyPtBi \cite{chen2022unconventional}.
For magnetic insulators, the in-plane thermal Hall effect has been debated to be an evidence for the chiral Majorana edge mode \cite{kitaev2006anyons} in a Kitaev spin liquid candidate $\alpha$-\ce{RuCl3} \cite{kasahara2018majorana,yokoi2021half,czajka2022planar,lefranccois2022evidence}.

Importantly, the in-plane Hall effect is a consequence of the absence of certain crystal symmetries \cite{juretschke1955symmetry,akgoz1975space,grimmer1993general}.
This is different from the situation for the normal Hall effect, which is always allowed by the time-reversal symmetry breaking under the out-of-plane field.
As a result, several factors need to be carefully considered in order to discuss the in-plane Hall effect.
In the view of the growing importance of this unconventional effect, we discuss the symmetry conditions and appropriate experimental protocols for the in-plane Hall effect, which provide a useful guide for the interpretation of the observed signal to extract the intrinsic feature of this phenomenon.

In this manuscript, we start by summarizing the symmetry requirements for the in-plane (thermal) Hall effect.
We derive the absence conditions for the in-plane Hall effect by applying a pictorial approach \cite{de1980pictorial}, which is an extension of the previous studies to arbitrary magnetic fields in nonmagnetic and magnetic materials.
It is applied to \ce{ZrTe5} to see if the observation in Ref. [\onlinecite{liang2018anomalous}] is allowed by symmetry.

The following three sections address the consequences of the lack of symmetries in the in-plane Hall system.
First, we consider the effect of crystal twinning, which degrades the observed in-plane Hall signal.
Second, we discuss the experimental protocol for estimating the in-plane Hall conductivity by the conversion between the (thermal) resistivity tensor $\hat{\rho}$ ($\hat{\lambda}$) and the conductivity tensor $\hat{\sigma}$ ($\hat{\kappa}$).
We show that the in-plane Hall resistivity in a (twin-free) three-dimensional system is contaminated by the out-of-plane transport, and is not directly proportional to the intrinsic in-plane Hall conductivity.
Interestingly, this effect is unavoidable due to the lack of symmetry for the in-plane Hall system.
Third, we consider the quasi-two-dimensional system, where the above effect is negligible.
Even in this case, the effect of the in-plane magnetic field breaks the $xy$-rotational symmetry, which restricts the application of the conversion formula from $\hat{\rho}$ ($\hat{\lambda}$) to $\hat{\sigma}$ ($\hat{\kappa}$), and requires an approach beyond the conventional five-electrode method using a single device.

Finally, as a case study, we consider the thermal Hall effect in $\alpha$-\ce{RuCl3} to see how the above factors affect the observations.
We point out that the procedures used in previous work to quantify the thermal Hall conductivity potentially contain a systematic error.
This is due to the breaking of the $xy$-rotational symmetry by the in-plane field, which has not been sufficiently verified in previous studies.
We propose the improvements of the experimental protocols in order to clarify the half-integer quantization of $\kappa_{xy}$ in the spin-disordered state of $\alpha$-\ce{RuCl3}.

As a note on the terminology, we emphasize that the in-plane Hall effect differs from the conventional planar Hall effect with respect to the response to the $B$-field reversal.
In this manuscript, to follow the convention \cite{zhou2022heterodimensional} and to avoid confusion, we use the term in-plane (thermal) Hall effect to refer to the field-odd response and the planar Hall effect only for the field-even response.
However, a few papers use the term "planar Hall effect" \cite{battilomo2021anomalous,cullen2021generating,czajka2022planar,takeda2022planar} to refer to the field-odd in-plane (anomalous/thermal) Hall effect.
We do not follow this trend, as the term "planar Hall effect" has long been used to evoke the field-even effect, although we do not say that this wording is inapprpriate \footnote{The \textit{Hall} effect is often used to describe only field-odd transverse responses \cite{jan1957galvamomagnetic,casimir1945onsager,zyuzin2020plane}, but phenomenologically it is more generally defined as being independent of the field-reversal symmetry \cite{beer1963galvanomagnetic,kao1958phenomenological}}.

\section{The symmetry conditions of the in-plane Hall effect}
Previous studies provide symmetry conditions of the in-plane Hall effect for nonmagnetic systems \cite{akgoz1975space,kao1958phenomenological,smith1967electronic}, for magnetic systems \cite{grimmer1993general} upto the $B$-linear term, and for the Berry curvature terms in nonmagnetic systems \cite{ren2016quantum,wang2022theory}.
There is also a specific application to a honeycomb-lattice system \cite{utermohlen2021symmetry}.
To more intuitively capture the importance of crystal symmetry, we apply a pictorial approach \cite{de1980pictorial} to obtain the necessary conditions for the in-plane Hall effect.
This is useful for heuristically deriving the symmetry conditions of various phenomena such as nonreciprocal phenomena and multiferroicity \cite{szaller2013symmetry,cheong2019sos,kurumaji2020spiral} in both magnetic and nonmagnetic systems

We consider the equation, $E_y=\rho_{yx}(B_x,B_y,0)J_x$, which relates the electric field along the $y$ axis and the applied current along the $x$ axis under the in-plane magnetic field $\bm{B}=(B_x,B_y,0)$.
We separate it into a field-even planar Hall component, $E^{e}_{y}\propto \rho^{e}_{yx}$ and a field-odd component, $E^{o}_{y}\propto \rho^{o}_{yx}$, and consider only the latter, i.e., the in-plane Hall effect.
We note that the conditions for $\rho^{o}_{yx}=0$ are equivalent to those for $\sigma^{o}_{xy}=0$ because the form of the conductivity tensor is identical to that of the resistivity.
All of these arguments hold regardless of the microscopic mechanism, and are even applicable to the in-plane thermal Hall effect by replacing the $E_y$ with the temperature gradient $-\nabla_y T$, and the current $J_x$ with the heat current.

\begin{figure}[t]
	\includegraphics[width =  \columnwidth]{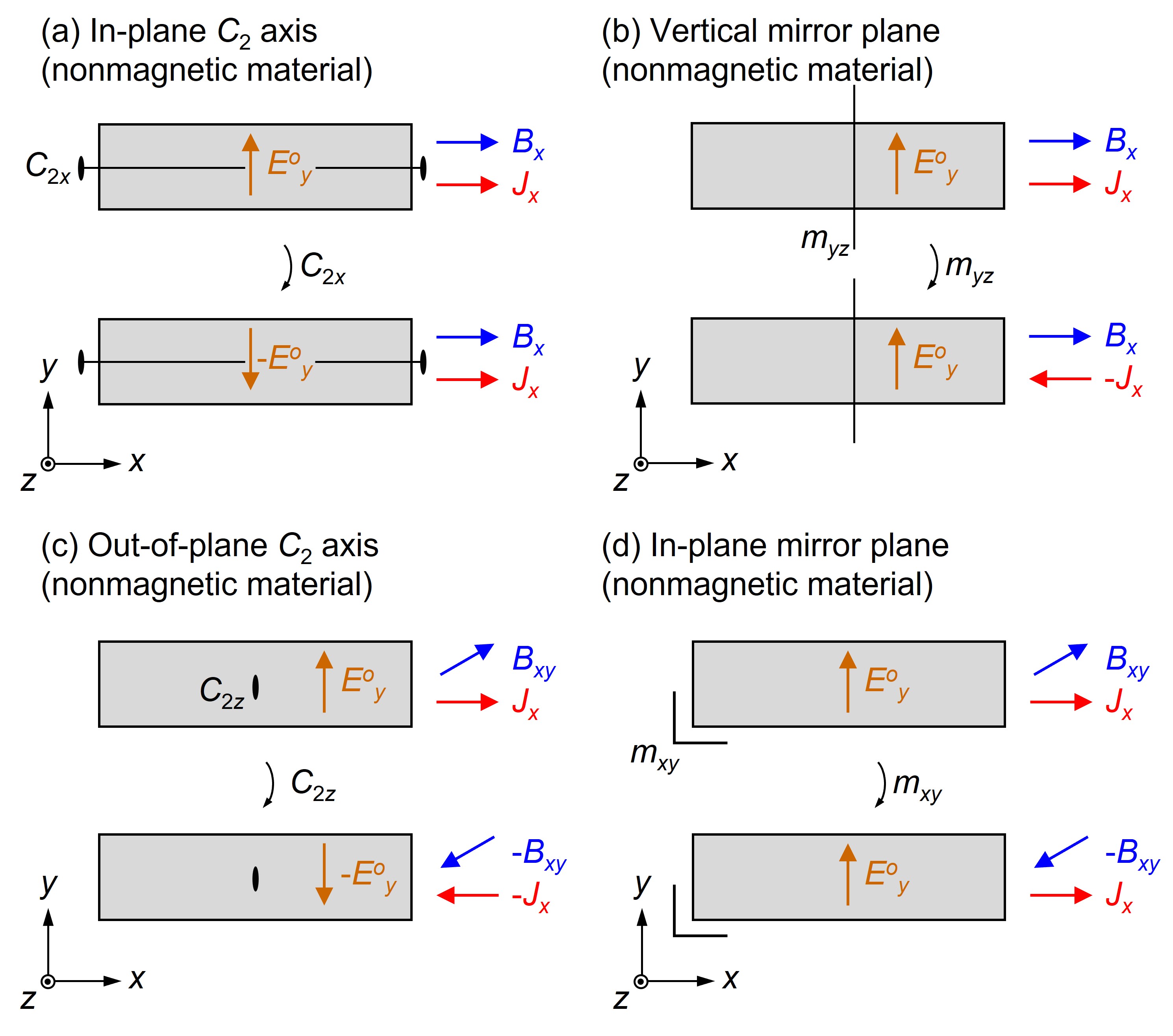}
	\caption{\label{figsymm}
 Symmetry conditions for the absence of the in-plane Hall effect.
 Gray rectangle is a crystal.
 Orange arrow represents a tentative in-plane Hall electric field $E^{o}_y$ proportional to $\rho^{o}_{yx}(B_x,B_y,0)$, which is proved to be zero.
 The superscript o denotes the field-odd nature.
 The rotation or mirror operation for each figure transforms the top panel into the bottom panel.
 The symmetry operations for the zero-field state are denoted as (a) a $C_2$ axis along the $x$ axis ($C_{2x}$, black line), (b) a vertical mirror in the $yz$ plane ($m_{yz}$, black line), (c) an out-of-plane $C_2$ axis along the $z$ axis ($C_{2z}$, black ellipse), and (d) a horizontal mirror in the $xy$ plane ($m_{xy}$, black L-shape symbol).
  }
\end{figure}

For simplicity, we consider nonmagnetic systems in the main text, and discuss the extension to magnetic materials in SI Sec. A.
We note that the in-plane Hall effect is allowed when the magnetic field is applied in an arbitrary direction with respect to the crystal axes \cite{akgoz1975space}, which reduces the crystal symmetry to the triclinic $1$ and $\bar{1}$.
We discuss rather special cases where the magnetic field is applied within a high-symmetry direction.
The following four cases are the symmetry conditions for the absence of the in-plane Hall effect.
\begin{enumerate}
\item $\rho_{yx}^{o}(B_x,0,0)$ is zero if there is a $C_2$ axis along the $x$ axis (Fig. \ref{figsymm}(a)).

\item $\rho^{o}_{yx}(B_x,0,0)$ is zero if there is a vertical mirror symmetry $m$ in the $yz$ plane (Fig. \ref{figsymm}(b)).

\item $\rho_{yx}^{o}(B_x,B_y,0)$ is zero if there is a $C_2$ axis along the $z$ axis (Fig. \ref{figsymm}(c)).

\item $\rho_{yx}^{o}(B_x,B_y,0)$ is zero if there is a horizontal mirror symmetry $m$ in the $xy$ plane (Fig. \ref{figsymm}(d)).

\end{enumerate}

Figures \ref{figsymm} provide how these above conditions are validated.
The top panel for each figure can be converted to the bottom one through the symmetry operation, which proves that $\rho^{o}_{yx}$.
We consider the condition 1 as an example.
Figure \ref{figsymm}(a) considers that the sample has the $C_2$ axis along the $x$ axis and current flows along the $x$ axis under a magnetic field parallel to $x$ axis.
We assume that the crystal has an intrinsic symmetry $C_2$ along the $x$ axis in zero field, and the applied field along $x$ does not break the symmetry (e.g., field-induced nematic/CDW/SDW transition involving the $C_2$ symmetry breaking).
This last assumption is necessary to guarantee that the crystal is identical when the $C_2$ rotation is applied even in a finite field.
As shown in the top panel of Fig. \ref{figsymm}(a), we assume a transverse electric field along the $y$ axis ($E^{o}_y$) due to the in-plane Hall effect.
We note that the experimental configuration is unchanged except for the direction of $E^{o}_y$, when we apply the $C_2$ rotation to the whole experimental setup including the $B$ and the applied $J$ (see the lower panel in Fig. \ref{figsymm}(a)).
This leads to the conclusion that $E^{o}_y = 0$, i.e., $\rho^{o}_{yx} = 0$.
In the same way, we can prove the other conditions 2-4 in SI Sec. A.

Here, we apply the above symmetry conditions to selected examples to see if the in-plane Hall effect is allowed.
Interesting contrast is obtained between the in-plane Hall effect in \ce{VS2}-\ce{VS}-superlattice \cite{zhou2022heterodimensional} and in \ce{ZrTe5} \cite{liang2018anomalous}.
As carefully discussed in Ref. [\onlinecite{zhou2022heterodimensional}], monoclinic unit cell allows the in-plane Hall effect, which is consistent with the experimental results.
In the case of an orthorhombic point group $mmm$, however, it is indeed forbidden as long as the field is in one of the mirror planes, i.e., $\rho^{o}_{xy}(B_x,B_y,0)=\rho^{o}_{yz}(0,B_y,B_z)=\rho^{o}_{zx}(B_x,0,B_z)=0$.
Accordingly, the reported in-plane Hall effect ($\rho_{zx}(B_x, 0, B_z)$) in \ce{ZrTe5} (space group: No. 63, $Cmcm$ at RT) \cite{liang2018anomalous} is forbidden by the $ac$ mirror symmetry, where the $xyz$ axes correspond to the crystallographic $abc$ axes.
The discrepancy between the symmetry condition and the experiments suggests an unrecognized symmetry lowering in the sample used, such as a monoclinic at low temperatures, or an unexpectedly large sensitivity to external shear strain or sample misalignment.

\section{Effect of lack of symmetry 1: twins}
\begin{figure}[t]
	\includegraphics[width =  0.8\columnwidth]{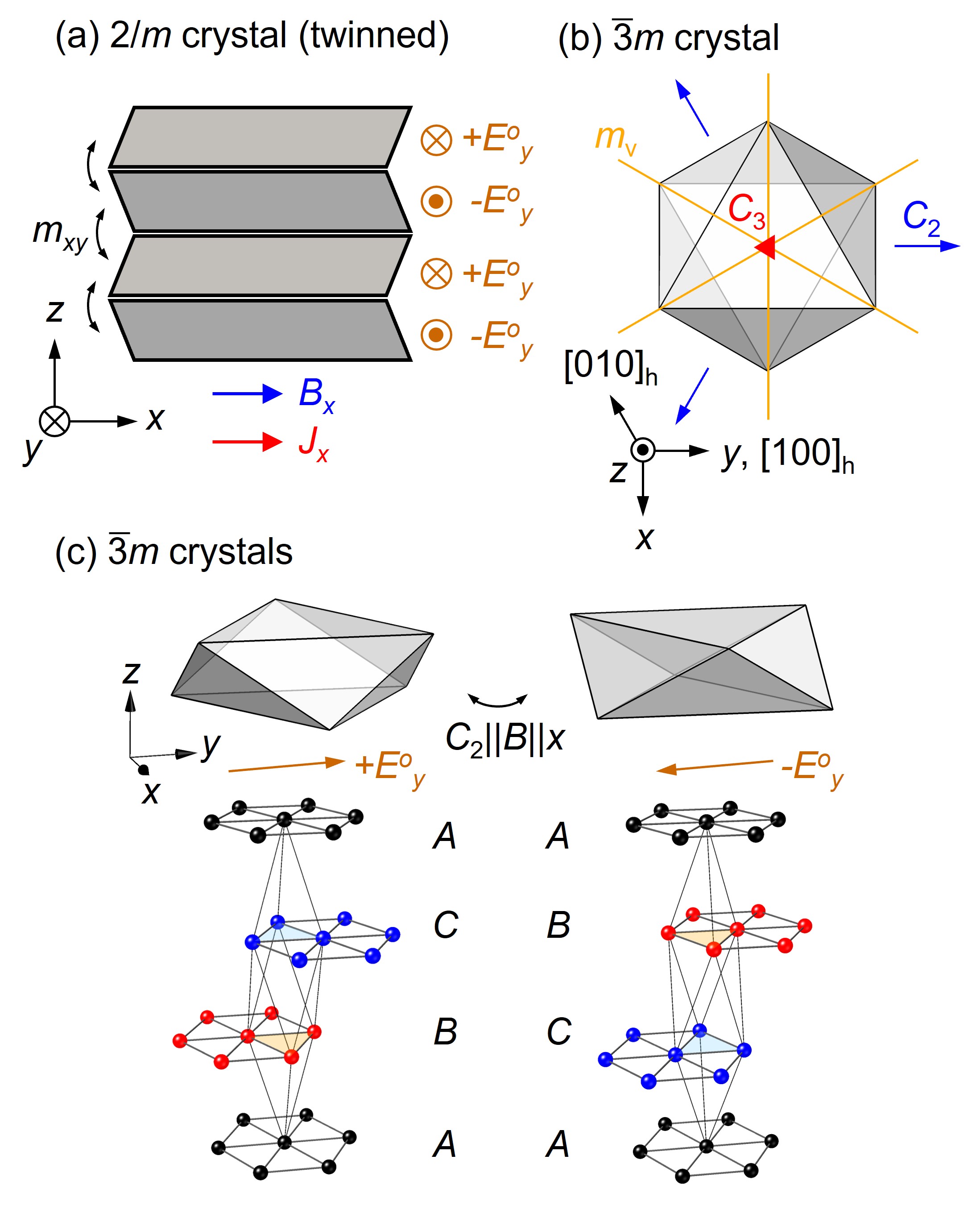}
	\caption{\label{figtwin}
 (a) Twinning by pseudo merohedry in terms of $m_{xy}$ for a $2/m$ crystal (the $C_2$ symmetry is along the $y$ axis).
 Gray parallelograms are twin domains, where the in-plane Hall effect ($\propto E^{o}_y$) is opposite. 
 (b) Schematic top-view of a trigonal crystal belonging to $\bar{3}m$ point group and symmetry elements.
 Red triangle: three-fold rotation along the $z$ axis, $C_3$.
 Blue arrows: in-plane two-fold rotations, $C_2$, where one of them is along the $y$ axis.
 There is no $C_2$ axis along the $x$ axis.
 Orange lines: vertical mirror planes, $m_{\text{v}}$, one of which is parallel to the $xz$ plane.
 Hexagonal axes ($[100]_{\text{h}}$ and $[010]_{\text{h}}$), and Cartesian coordinates, $x$, $y$, and $z$ are also shown.
 (c) Two $\bar{3}m$ crystals with respect to the twinning by reticular merohedry (obverse-reverse twins), where the $C_2$ rotation along the $x$ axis transforms one from the other.
  The induced in-plane Hall electric field $E^{o}_y$ is reversed.
 The bottom panel shows a corresponding schematic rhombohedral $ABC$ and $CBA$-stacking of triangular lattices, where a unit cell is emphasized by black lines.
  }
\end{figure}

As shown above, the in-plane Hall effect requires that the crystals lack certain symmetries.
In real materials, this feature potentially causes twinning to cancel the signal expected in a monodomain.
In practice, twinned crystals are avoided for measurements, but sometimes careful inspection miss a twin by merohedry \cite{parsons2003introduction}, where its twin operation belongs to the holohedry point group (higher-symmetry group of the crystals).
When the twin operation reverses the in-plane Hall voltage, the twin domains, crystal 1 and 2, contribute oppositely to the signal.
If $v$ is the volume ratio of the crystal 1, the total in-plane Hall signal is
\begin{equation}
\rho^{o,\text{tot}}_{yx}(B_x,0,0)=v\rho^{o,1}_{yx}+(1-v)\rho^{o,2}_{yx}=(2v-1)\rho^{o}_{yx},
\end{equation}
which vanishes at $v=0.5$.
Similar situation occurs due to a twin by pseudo-merohedry \cite{parsons2003introduction} in a monoclinic crystal as shown in Fig. \ref{figtwin}(a).
Twinning due to a lack of mirror symmetry parallel to the $xy$ plane results in the opposite contribution from each domain.

Another case is for a layered rhombohedral crystal, which potentially contains twinning by reticular merohedry (obverse-reverse twinning) \cite{parsons2003introduction}.
We consider the crystal belonging to the point group $\bar{3}m$ (Fig. \ref{figtwin}(b)).
As the $C_2$ axis is absent along the $x$ axis, the crystal often twinned with the domain associated with the $C_2$ rotation along the $x$ axis (Fig. \ref{figtwin}(c), top).
Each domain contributes oppositely to the in-plane Hall signal along the $y$ axis under $B_x$.
This twinning occurs for example when there is the stacking fault regarding the inversion of $ABC$-stacked layers to $CBA$-stacked layers of triangular lattices (Fig. \ref{figtwin}(c), bottom).
To accurately evaluate the magnitude of the intrinsic in-plane Hall effect, we need to select de-twinned crystals, e.g., by checking crystallographic morphology or the extinction rules of the diffraction patterns.

\section{Effect of lack of symmetry 2: Out-of-plane transport}
We consider a protocol for evaluating the intrinsic in-plane Hall conductivity for a twin-free crystal.
We start by considering the relationship between conductivity and resistivity.
A conductivity tensor ($\hat{\sigma}$) is a $3\times 3$ matrix connecting an applied electric field ($\bm{E}$) and an induced current ($\bm{J}$):
\begin{equation}
\bm{J}=\hat{\sigma}\bm{E}.
\end{equation}
Instead of measuring the conductivities directly, one usually attempts to measure resistivities because it is much easier to control the direction of the $\bm{J}$ in a sample than to manage the $\bm{E}$.
The resistivity tensor $\hat{\rho}$ connects the $\bm{J}$ and $\bm{E}$ inversely from the $\hat{\sigma}$:
\begin{equation}
\bm{E}=\hat{\rho}\bm{J}
\end{equation}
As the theories usually provide the predictions of the conductivity, the accurate conversion of the experimental observation $\hat{\rho}$ to the $\hat{\sigma}$ is important to identify the underlying mechanisms of the in-plane Hall effect. 

Here, we consider the following question.
Suppose that we observe a non-zero $\rho^{o}_{yx}(B_x)$, can we immediately conclude that the signal is exclusively ascribed to the finite in-plane Hall conductivity ($\sigma^{o}_{xy}$)?
It is, in fact, not the case because $\rho^{o}_{yx}(B_x)$ is contaminated with the effect of the out-of-plane transport in the three-dimensional system.

\begin{figure}[t]
	\includegraphics[width =  0.8\columnwidth]{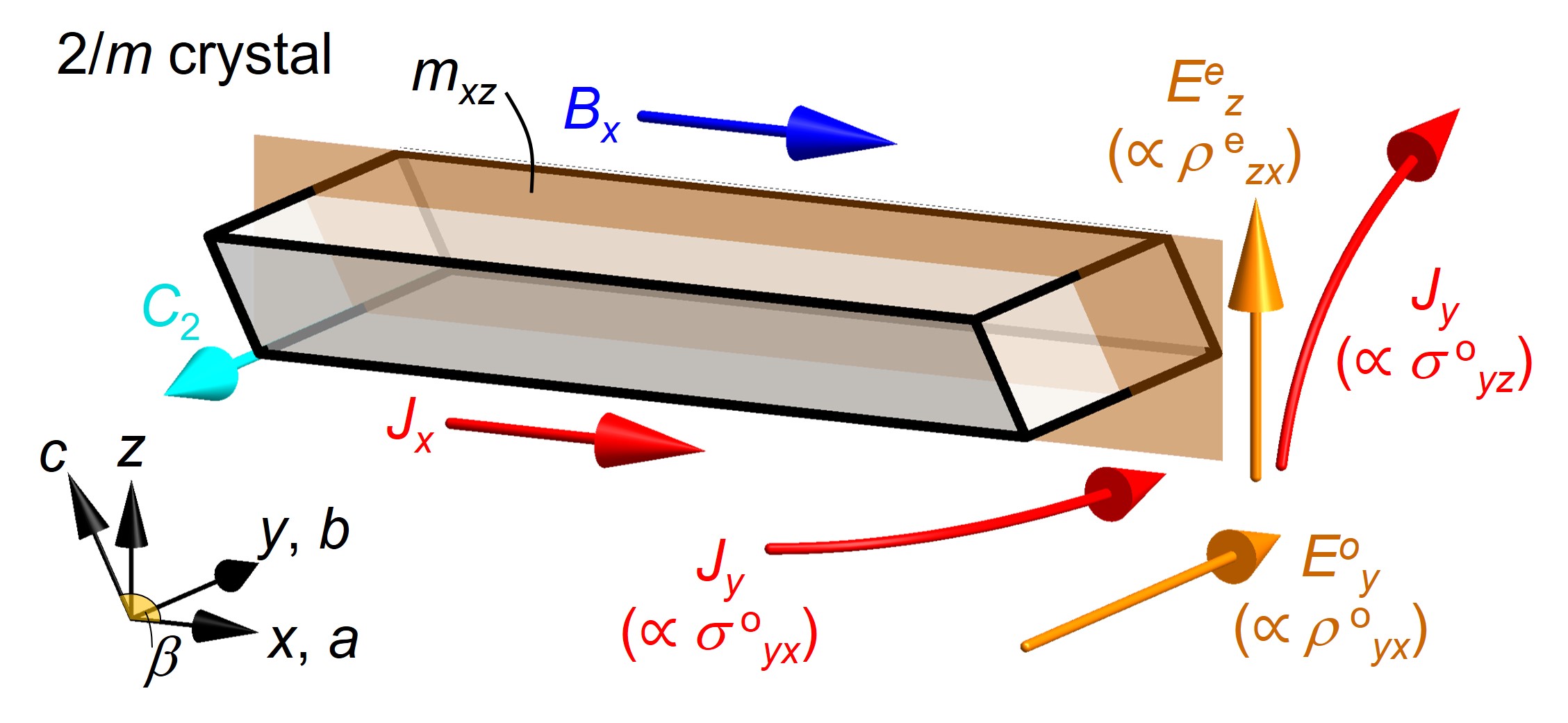}
	\caption{\label{fig2/m}
Schematics of a $2/m$ crystal (gray parallelepiped) and the in-plane Hall effect.
 Orange plane is the mirror ($m_{xz}$) parallel to the $xz$ plane.
 Cyan arrow is the $C_2$ axis along the $y$ axis.
 $xyz$ is the Cartesian coordinate introduced as $x\parallel a$ and $y\parallel b$, and $\beta$ ($>90^{\circ}$) is the angle between the $a$ and $c$ axes.
 The in-plane Hall effect $E_y$ ($\propto \rho^{o}_{yx}$, oblique orange arrow) is allowed under $J\parallel B\parallel x$.
 The out-of-plane electric field $E_z$ (vertical orange arrow) is induced by the crystalline planar Hall coefficient ($\propto \rho^{e}_{zx}J_x$).
 The out-of-plane current is deflected towards $y$ by the Lorentz force from the $B_x$ to induce $J_y$ ($\propto \sigma^{o}_{yz}E_z$, red arrow on the right), which is added to the intrinsic $J_y$ ($\propto \sigma^{o}_{yx}$, red arrow in the middle).
  }
\end{figure}

We consider a monoclinic $2/m$ system (Fig. \ref{fig2/m}), which provides a typical formula of our interest with respect to the in-plane Hall conductivity.
The conductivity tensor is as follows,
\begin{equation}\label{sigma2/m}
\hat{\sigma}_{2/m}(B_{x})=
\begin{pmatrix}
\sigma^{e}_{xx} & \sigma^{o}_{xy} & \sigma^{e}_{xz}\\
-\sigma^{o}_{xy} & \sigma^{e}_{yy} & \sigma^{o}_{yz}\\
\sigma^{e}_{xz} & -\sigma^{o}_{yz} & \sigma^{e}_{zz}
\end{pmatrix}
.
\end{equation}
Here, we set the $C_2$ axis parallel to the $y$ axis and the mirror plane in the $xz$ plane (Fig. \ref{fig2/m}).
The $z$ axis is defined as orthogonal to the $xy$ plane.
The superscripts, e and o, are for field-even and field-odd quantities, respectively.
Since the tensor form of $\hat{\rho}$ is identical with that of $\hat{\sigma}$, we put
\begin{equation}\label{rho2/m}
\hat{\rho}_{2/m}(B_{x})=
\begin{pmatrix}
\rho^{e}_{xx} & -\rho^{o}_{yx} & \rho^{e}_{zx}\\
\rho^{o}_{yx} & \rho^{e}_{yy} & -\rho^{o}_{zy}\\
\rho^{e}_{zx} & \rho^{o}_{zy} & \rho^{e}_{zz}
\end{pmatrix}
.
\end{equation}

By taking the inverse matrix, $\hat{\rho}=\hat{\sigma}^{-1}$, we obtain the formula for $\rho^{o}_{yx}$:
\begin{equation}\label{outofplaneryx}
\rho^{o}_{yx}=\frac{\sigma^{e}_{zz}}{\Delta_{\sigma}}(\sigma^{o}_{xy}+\sigma^{e}_{xz}\sigma^{o}_{yz}/\sigma^{e}_{zz}),
\end{equation}
where $\Delta_{\sigma}$ is the determinant of $\hat{\sigma}$ (see SI Sec. B).
We note that the $\rho^{o}_{yx}$ is not directly proportional to the in-plane Hall conductivity $\sigma^{o}_{xy}$, but is contaminated by the term $\sigma^{e}_{xz}\sigma^{o}_{yz}/\sigma^{e}_{zz}$, where $\sigma^{o}_{yz}$ corresponds to the Hall effect in the $yz$ plane under $B_x$.
This term is finite if the field-even component $\sigma^{e}_{xz}$ is finite.
We refer $\sigma^{e}_{xz}$ (and $\rho^{e}_{zx}$) to the crystalline planar Hall effect (see SI Sec. C).

The contamination of the cross term between the $\sigma^{o}_{yz}$ and $\sigma^{e}_{xz}$ can be understood as shown in Fig. \ref{fig2/m}.
The application of the current $J_x$ generates the out-of-plane electric field $E_z$ due to $\rho^{e}_{zx}$, which leads to the out-of-plane conduction $J_z$.
In the presence of the $B_x$, the $J_z$ is deflected in the $yz$ plane to produce the transverse $J_y\propto \sigma ^{o}_{yz}$, which produces an $E_y$ together with the intrinsic in-plane Hall current ($J_y\propto \sigma ^{o}_{yx}$) \footnote{In a single-carrier system, it can be proved that those two terms in Eq. (\ref{outofplaneryx}) completely cancel with each other \cite{bauhofer1985longitudinal} unless the Berry curvature has finite contribution. In two-carrier system, this cancellation is imperfect due to mobility difference.}.

It is evident that in a monoclinic crystal the $\sigma^{e}_{xz}$ can be finite even in zero field because $\beta \neq 90^{\circ}$ does not guarantee the diagonalized form of the conductivity tensor.
As a matter of fact, the effect due to $\sigma^{e}_{xz}$ is unavoidable for higher-symmetry systems when we consider the in-plane Hall effect of a three-dimensional system at finite field.
We can prove that the symmetry condition for the finite in-plane Hall effect $\sigma^{o}_{xy}$ is equivalent to that for the $\sigma^{e}_{xz}$ (see SI Sec. C).

This contamination effect is serious if there is a significant anomalous (Berry curvature) contribution in $\sigma^{o}_{yz}$.
We consider that Hall conductivity can be divided into two terms, normal (Lorentz force origin) and anomalous (Berry curvature origin) components, i.e., $\sigma^{o}_{xy}=\sigma^{N}_{xy}+\sigma^{A}_{xy}$ and $\sigma^{o}_{yz}=\sigma^{N}_{yz}+\sigma^{A}_{yz}$.
The anomalous in-plane Hall resistivity $\rho^{A}_{yx}(B_x)$ is, then, expressed as follows.
\begin{equation}\label{ryxAN}
\rho^{A}_{yx}=\frac{\sigma_{zz}}{\Delta_{\sigma}}(\sigma^{A}_{xy}+\sigma_{xz}\sigma^{A}_{yz}/\sigma_{zz}).
\end{equation}
Here, we omit the superscripts, e and o, for simplicity.
Evidently, we cannot directly connect the observation of the $\rho^{A}_{yx}$ with the presence of the $\sigma^{A}_{xy}$, and need to exclude the effect from $\sigma^{A}_{yz}$.

For quasi-two-dimensional systems, the situation is expected to become simpler.
We assume that $\sigma^{A}_{yz}$ is negligible.
The $yz$ components is small because we can prove that $\sigma^{N}_{yz}\propto \mu_{z}$, where $\mu_{z}$ is the out-of-plane carrier mobility.
As the magnitude of $\sigma^{e}_{xz}$ cannot exceeds $\sqrt{\sigma^{e}_{xx}\sigma^{e}_{zz}}$ (since $\Delta_{\sigma}>0$), we can reasonably expect that the $xz$ component is also restricted by the low $\mu_{z}$ ($\propto \sigma^{e}_{zz}$).
These features can validate the assumption for ignoring the effect of the out-of-plane transport in the layered systems, and the approximation taking the resistivity and conductivity tenors as $2\times 2$ matrices.
In the next section, we consider the case of the quasi-two-dimensional system and how the $\sigma^{o}_{xy}$ is measured.

\section{Effect of lack of symmetry 3: absence of $xy$-rotational symmetry.}
In this section, we ignore the $xz$ and $yz$ components in $\hat{\sigma}$ and $\hat{\rho}$ (Eqs. (\ref{sigma2/m}), (\ref{rho2/m})) for the quasi-two-dimensional system.
Even in such a simplified situation, we have to consider the absence of the $xy$-rotational symmetry.

\begin{figure}[t]
	\includegraphics[width = \columnwidth]{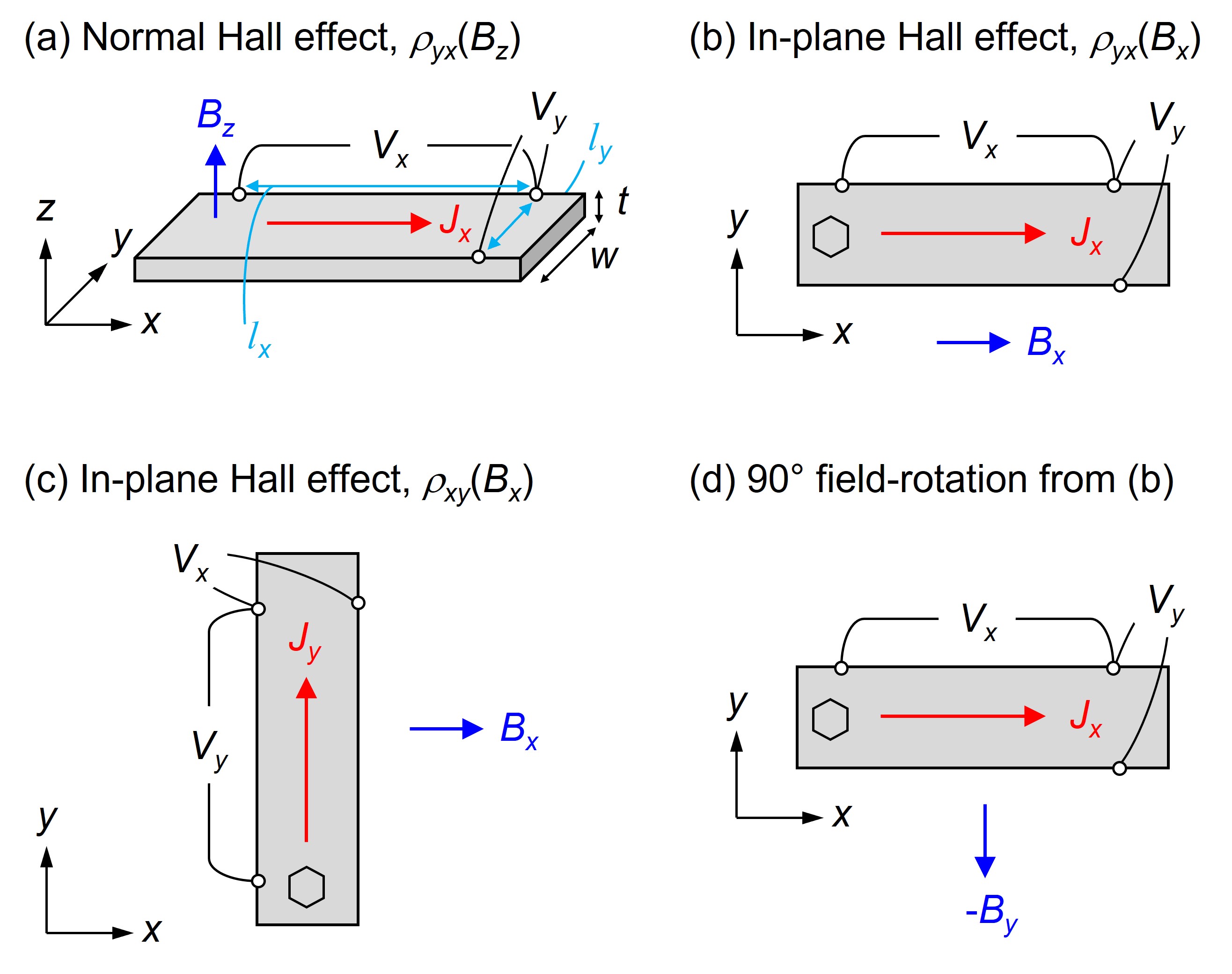}
	\caption{\label{figFive}
 Schematic five-electrode configurations for measurements of (a) a normal Hall effect, and (b) an in-plane Hall effect.
 The gray rectangle is a shaped sample with width $w$ and thickness $t$.
 White circles are electrodes, and black lines are wires to measure longitudinal voltage drop $V_x$ and transverse voltage difference $V_y$ in an applied current along the $x$ axis, $J_x$.
 The magnetic field is applied (a) perpendicular to the $xy$ plane, $B_z$, and (b) along the $x$ axis, $B_x$.
 In the setup of (b), we obtain $\rho_{xx}(B_x)$ and $\rho_{yx}(B_x)$.
 (c) is another setup for the in-plane Hall effect with $J_y$ providing $\rho_{yy}(B_x)$ and $\rho_{xy}(B_x)$.
 (d) is the setup with $90^{\circ}$ rotation of the magnetic field from (b), giving $\rho_{xx}(-B_y)$ and $\rho_{yx}(-B_y)$
 A hexagon marker on each corner denotes the in-plane orientation of the sample (see text).
 }
\end{figure}

To emphasize the distinction from the conventional normal Hall effect, we first consider the measurement of the isotropic system under a magnetic field along the $z$ axis (Fig. \ref{figFive}(a)).
A five-electrode measurement using a single-device is sufficient to estimate $\sigma^{o}_{xy}$, where two electrodes are connected to a current source, and three are used to monitor longitudinal and transverse voltage drops.
We obtain $\rho^{e}_{xx}(B_z)$ $(=V_x/J_x\cdot wt/l_x)$ and $\rho^{o}_{yx}(B_z)$ $(=V_y/J_x\cdot wt/l_y)$, where $l_x$ and $l_y$ are the electrode distance for $V_x$ and $V_y$, respectively. 
From the view-point of symmetry and Onsager's relation, we can put $\rho^{e}_{yy}(B_z)=\rho^{e}_{xx}(B_z)$ and $\rho^{o}_{xy}(B_z)=-\rho^{o}_{yx}(B_z)$.
We obtain the single-device formula to estimate the normal Hall conductivity
\begin{equation}\label{NHE}
\sigma^{o}_{xy}(B_z)=\rho^{o}_{yx}/[(\rho^{e}_{xx})^2+(\rho^{o}_{yx})^2].
\end{equation}

In contrast to the above case, the $\sigma^{o}_{xy}$ in the in-plane Hall system cannot be estimated within a single setup of the five-electrode measurement (Fig. \ref{figFive}(b)).
The magnetoresistivity tensor when the magnetic field is applied along the $x$ axis is given by
\begin{equation}\label{rho2D}
\hat{\rho}_{2D}(B_x)=
\begin{pmatrix}
\rho^{e}_{xx} & -\rho^{o}_{yx}\\
\rho^{o}_{yx} & \rho^{e}_{yy}
\end{pmatrix}
.
\end{equation}
Here, we assume that the symmetry of the system is high enough to have the purely field-odd off-diagonal component, i.e., no planar Hall effect (this is not true, for example, for $3$ and $\bar{3}$ systems \cite{akgoz1975space}).
Importantly, the inequality $\rho^{e} _{yy}(B_x) \neq \rho^{e} _{xx}(B_x)$ is unavoidable, in principle, because the in-plane magnetic field $B_x$ always breaks the rotational symmetry around the $z$ axis even in the case of the high-symmetric system in zero field (e.g., $\bar{3}m$, $3m$, and $32$).
The origin of the anisotropy can be easily understood because the longitudinal current ($J_x\parallel B_x$) is less affected by the Lorentz force than the transverse current ($J_y \perp B_x$).
We also note that, in fact, the inequality $\rho^{e}_{xx} \neq \rho^{e}_{yy}$ often occurs in the in-plane Hall systems because it is mainly allowed in low-symmetric systems such as monoclinic crystals \cite{mollendorf1984first,bauhofer1988two,zhou2022heterodimensional}.

The inverse matrix of the resistivity tensor gives the longitudinal conductivity $\sigma^{e}_{xx}$ as follows.
\begin{multline}\label{sigmaxx}
\sigma^{e} _{xx}(B_x)=\rho^{e}_{yy}/[\rho^{e}_{xx}\rho^{e}_{yy}+(\rho^{o}_{yx})^2]\\
\sim \rho^{e}_{yy}/(\rho^{e}_{xx}\rho^{e}_{yy})\sim 1/\rho^{e}_{xx}
\end{multline}
Here, we assume in the last equality that the Hall angle ($\rho^{o}_{yx}/\rho^{e}_{xx}$) is negligible.
We find that the longitudinal conductivity can be estimated with sufficient accuracy within a five-electrode measurement of the setup in Fig. \ref{figFive}(b).
The Hall conductivity ($\sigma^{o}_{xy}$), on the other hand, is obtained in the exact form by using all three independent quantities, $\rho^{e}_{xx}$, $\rho^{o}_{yx}$, and $\rho^{e}_{yy}$ (see SI Sec. B for the effect of three-dimensionality):
\begin{equation}\label{sigmaxyprecise}
\sigma^{o,\text{exact}} _{xy}(B_x)=\rho^{o}_{yx}/[\rho^{e}_{xx}\rho^{e}_{yy}+(\rho^{o}_{yx})^2]\sim \rho^{o}_{yx}/(\rho^{e}_{xx}\rho^{e}_{yy}).
\end{equation}
To estimate $\rho^{e}_{yy}(B_x)$, we need to prepare another sample as shown in Fig. \ref{figFive}(c) with a long edge along the $y$ axis and perform an experiment for $J_y$ under $B_x$.
We cannot measure $\rho^{e}_{yy}(B_x)$ with the same sample for the $J_x\parallel B_x$ (Fig. \ref{figFive}(b)) because it has to be shaped into a thin rectangular parallelpiped with a long edge along the $x$ axis to increase the signal ($\propto l/wt$) as well as to suppress the geometrical effect \cite{jan1957galvamomagnetic,isenberg1948improved,drabble1957geometrical,mumford2020sample}.

One might consider that we can measure $\rho^{e}_{xx}(-B_y)$ without changing the sample by rotating the magnetic field from $B_x$ to $-B_y$ (Fig. \ref{figFive}(d)), which seems to be an alternative to $\rho^{e}_{yy}(B_x)$ in Eq. (\ref{rho2D}).
This is in fact inapplicable because $\rho^{e}_{xx}(-B_y) \neq \rho^{e}_{yy}(B_x)$, which can be seen from the fact that a hexagon symbol on the sample in Fig. \ref{figFive}(d) changes its orientation when one tries to superimpose it on Fig. \ref{figFive}(c) by rotating the whole setup by $-90^{\circ}$.
The inequality $\rho^{e}_{xx}(-B_y) \neq \rho^{e}_{yy}(B_x)$ arises from the symmetry condition that forbids a $C_{2z}$ symmetry for the non-zero in-plane Hall effect  (see Fig. \ref{figsymm}(c)), because this rule forbids $C_{4z}$ symmetry at the same time.
In other words, if there is a symmetry that ensures $\rho^{e}_{xx}(-B_y)=\rho^{e}_{yy}(B_x)$, one can prove $\rho^{o}_{yx}(B_x)=0$, which is not the situation considered here.

Since the $\rho^{e}_{yy}(B_x)$ cannot be obtained within the single-device measurement using the setup in Fig. \ref{figFive}(b), the above formula is often replaced by the following approximate form: 
\begin{equation}\label{sigmaxypseudo}
\sigma^{o, \text{approx}} _{xy}(B_x)=\rho^{o}_{yx}/[(\rho^{e}_{xx})^2+(\rho^{e}_{yx})^2]\sim \rho^{o}_{yx}/(\rho^{e}_{xx})^2
\end{equation}
This prescription resembles Eq. (\ref{NHE}) for the normal Hall effect, but contains a leap of logic implicitly assuming that $\rho^{e}_{xx}(B_x)=\rho^{e}_{yy}(B_x)$, which is in general invalid for the in-plane Hall system.
Nevertheless, this approximate form of the Hall conductivity has often been used in many related experiments for orthorhombic/monoclinic compounds \cite{zhou2022heterodimensional,gourgout2022magnetic,fujioka2019strong,ge2020unconventional}, tilted-field configuration \cite{nayak2016large,hirschberger2021nanometric,kasahara2018majorana}, $xz$-plane-transport of tetragonal systems \cite{alam2022sign}, as well as for the in-plane thermal Hall conductivity in magnetic insulators \cite{yokoi2021half,czajka2022planar, takeda2022planar}.

Although this does not bring a big problem in many cases for the estimation of the order of magnitude, it is crucial when the accurate value of the (thermal) Hall conductivity is connected with the theoretical interpretation of the quantum nature of the material.
In principle, we need to prepare two independent devices (Figs. \ref{figFive}(b)-(c)) and measure $\rho^{e}_{xx}$, $\rho^{e}_{yy}$, and $\rho^{o}_{yx}$, to apply Eq. (\ref{sigmaxyprecise}) with all three independent components. 
In the next section, we discuss the in-plane thermal Hall effect in $\alpha$-\ce{RuCl3}, where the quantitative evaluation of the half-integer quantization has been concluded on the basis of the approximate form of $\kappa^{o}_{xy}$.

\section{Case of the half-integer quantization of thermal Hall effect in $\alpha$-\ce{RuCl3}}
Evidently, all of the above arguments are applicable to the in-plane thermal Hall effect in insulators.
In particular, an accurate evaluation of the thermal Hall conductivity is important as argued in recent studies on the half-integer quantization of $\kappa_{xy}$ under an in-plane and a tilted magnetic field in a Kitaev spin-liquid candidate $\alpha$-\ce{RuCl3} \cite{kasahara2018majorana,yokoi2021half,bruin2022robustness,yamashita2020sample,kasahara2022quantized}.

\begin{figure}[t]
	\includegraphics[width =  \columnwidth]{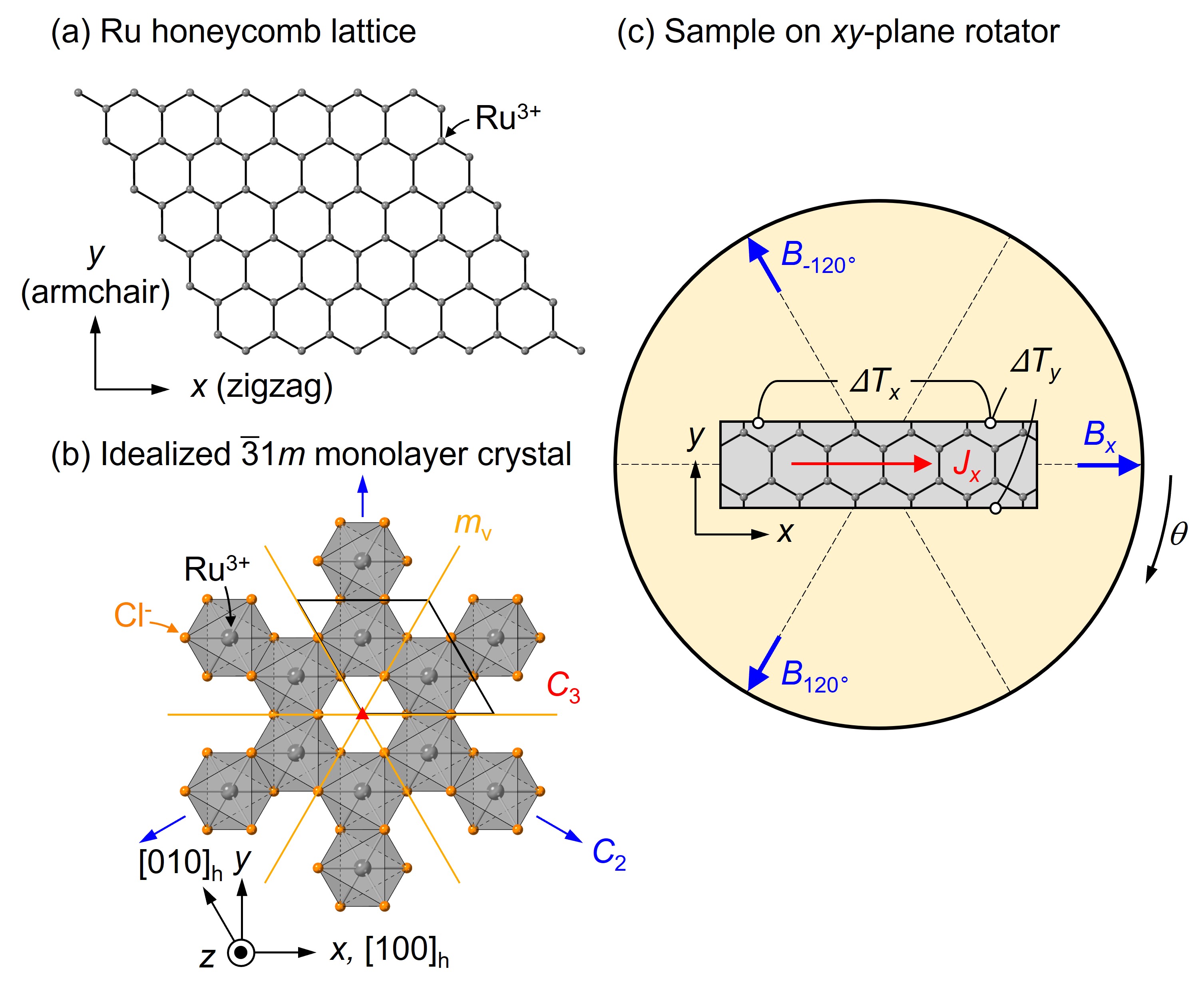}
	\caption{\label{fig3barm1} (a) Schematic honeycomb lattice of Ru$^{3+}$ ions.
 Zigzag ($\parallel x$) and armchair ($\parallel y$) axes are defined.
 (b) The idealized honeycomb layer of $\alpha$-\ce{RuCl3} belonging to the point group $\bar{3}1m$.
 Hexagonal axes, Cartesian coordinates, $x$, $y$, and $z$, the unit cell (black rhombus), and symmetry elements are also shown (see Fig. \ref{figtwin}(b)).
 (c) Schematic improved single-device setup for the in-plane thermal Hall effect of $\alpha$-\ce{RuCl3} under the in-plane magnetic field along $x$ axis ($B_x$), and directions rotated by $\pm 120^{\circ}$ around the $z$ axis ($B_{\pm120^{\circ}}$).
 The electrodes are for the measurement of longitudinal ($\Delta T_x$) and transverse ($\Delta T_y$) temperature differences.
}
\end{figure}

The crystal structure of $\alpha$-\ce{RuCl3} consists of a stacked honeycomb-lattice of Ru$^{3+}$ with magnetic moments (Fig. \ref{fig3barm1}).
There are two orthogonal axes in the lattice plane, which are called zigzag ($x$ axis) and armchair ($y$ axis) directions, perpendicular and parallel to a Ru-Ru bond, respectively.
$\alpha$-\ce{RuCl3} is antiferromagnetically ordered at below 7 K, and an in-plane magnetic field of 7 T is required to suppress the ordered state \cite{kasahara2018majorana}.

The half-integer quantization of the thermal Hall conductivity is observed in the field-induced spin-disordered state when the heat current is applied along the zigzag direction ($x$ axis) and the magnetic field is applied along the same direction or tilted from the axis perpendicular to the honeycomb-lattice plane ($z$ axis) towards the $x$ axis \cite{kasahara2018majorana,yokoi2021half,bruin2022robustness,yamashita2020sample,kasahara2022quantized}.
The in-plane Hall effect in a honeycomb layer is allowed by symmetry \cite{utermohlen2021symmetry}, and the absence of thermal Hall effect in the magnetic field along the armchair direction ($y$ axis) \cite{yokoi2021half} follows the symmetry condition of honeycomb-lattice of edge-shared \ce{RuCl6} octahedra \cite{utermohlen2021symmetry}.
The $\kappa_{xy}$ is normalized to each honeycomb-lattice with the layer separation by $d$: $\kappa^{\text{2D}}_{xy}/T=\kappa_{xy}\cdot d/T$, which shows a quantization to the half value of $K_0=(\pi^2k^2_{\text{B}}/3h)T$, implying the presence of the edge current of Majorana fermions \cite{kitaev2006anyons,nasu2017thermal}.

Indeed, the access to the spin-disordered state requires the in-plane component of the magnetic field, which breaks the $xy$-plane rotational symmetry regardless of the crystal structure at low temperature.
The exact form of the in-plane thermal Hall conductivity is expressed by
\begin{equation}\label{kappaxyprecise}
\kappa^{o,\text{exact}}_{xy}(B_x)\sim \lambda^{o}_{yx}/\lambda^{e}_{xx}\lambda^{e}_{yy},
\end{equation}
and it requires all the three independent thermal resistivity components, $\lambda^{e}_{xx}(B_x)$, $\lambda^{e}_{yy}(B_x)$, and $\lambda^{o}_{yx}(B_x)$ (we can reasonably ignore the effect of the out-of-plane transport; see below).
In previous studies, however, the thermal Hall conductivity has been estimated by the approximate form \cite{kasahara2018majorana,kasahara2022quantized,yamashita2020sample,czajka2021oscillations,czajka2022planar,bruin2022origin}:
\begin{equation}\label{kappaxypseudo}
\kappa^{o,\text{approx}}_{xy}(B_x)\sim \lambda^{o}_{yx}/(\lambda^{e}_{xx})^2,
\end{equation}
where the equivalence $\lambda^{e}_{xx}(B_x)=\lambda^{e}_{yy}(B_x)$ is implicitly assumed.

To the best of our knowledge, the component $\lambda^{e}_{yy}(B_x)$ has scarcely been reported, i.e., the thermal conductivity with heat current along the armchair direction ($y$ axis) under a field along the zigzag direction ($x$ axis), except for Ref. [\onlinecite{lefranccois2023oscillations}].
We summarize the references that report the thermal transport properties of $\alpha$-\ce{RuCl3} and experimental conditions in SI Sec. D.
The thermal Hall experiment with $J_y$ under $B_x$ is missing.
As discussed in the above sections, we note that the $\lambda^{e}_{yy}(B_x)$ cannot be replaced by $\lambda^{e}_{xx}(-B_y)$, which is realized by rotating the in-plane field from $B_x$ to $B_y$ without changing the sample setup for $J\parallel x$.
It is also known from the experimental results giving the different magnetic phase diagrams for $B_x$ and $B_y$ \cite{bruin2022robustness,suetsugu2022evidence}.
Reference \cite{lefranccois2023oscillations} reports both $\lambda^{e}_{xx}(B_x)$ and $\lambda^{e}_{yy}(B_x)$ (but no $\lambda^{o}_{yx}(B_x)$) with different samples.
They show similar $B_x$-field dependence but their magnitudes are discernibly different, implying the possible anisotropy between $\lambda^{e}_{xx}(B_x)$ and $\lambda^{e}_{yy}(B_x)$.

Although the significant $\kappa^{o}_{xy}$ in $\alpha$-\ce{RuCl3} has been repeatedly observed and the plateau feature has been reproduced \cite{kasahara2018majorana,yamashita2020sample,bruin2022robustness,kasahara2022quantized}, there are still debates in terms of the accurate value in the spin-disordered region \cite{czajka2022planar,lefranccois2022evidence}, and microscopic origins of the $\kappa^{o}_{xy}$ \cite{czajka2022planar,chern2021sign,zhang2021topological,koyama2021field,lefranccois2022evidence}.
As the theoretical predictions are provided by the $\kappa^{o}_{xy}$, the accurate conversion process from the experimentally measured thermal resistivity $\hat{\lambda}$ to the thermal conductivity $\hat{\kappa}$ is important.
The controversy is potentially due to the estimation of $\kappa^{o}_{xy}$ by the application of the approximate form Eq. (\ref{kappaxypseudo}), which may be sensitive to the sample dependence of anisotropy between $\lambda^{e}_{xx}$ and $\lambda^{e}_{yy}$. 
For a more quantitative verification of the half-integer quantization, an experimental proof of the hypothesis $\lambda^{e}_{yy}(B_x)\simeq \lambda^{e}_{xx}(B_x)$ in the spin-disordered field region has to be provided at least within the precision of better than 10\% comparable to that of the geometric errors \cite{yokoi2021half}.
A practical difficulty would arise from the notorious sample dependence \cite{yamashita2020sample,bruin2022robustness}, which would obscure the possible equivalence between $\lambda^{e} _{yy}(B_x)$ and $\lambda^{e} _{xx}(B_x)$.
We discuss two experimental protocols that take into account the sample quality variations and can be applied to $\alpha$-\ce{RuCl3} to quantitatively verify the half-integer value of $\kappa_{xy}$. 

One of the approaches is to make use of the three-fold rotational symmetry, which is supposed to be realized in the low-temperature crystal structure of $\alpha$-\ce{RuCl3} \cite{tanaka2022thermodynamic,park2016emergence}, although it seems to be still under discussion \cite{bruin2022origin,kim2022alpha,lebert2022acoustic,cao2016low}.
The advantage of this method is that it can directly measure the anisotropy of the thermal resistivity $\lambda^{e}_{xx}-\lambda^{e}_{yy}$ under the magnetic field without changing the sample (improved single-device method).
The idea is similar to the method proposed in Refs. [\onlinecite{shapiro2016measurement,walmsley2017determination}], where the elastoresistive effect is considered.

Here, we assume that the crystal structure of a monolayer $\alpha$-\ce{RuCl3} effectively belongs to the $\bar{3}1m$ point group, as shown in Fig. \ref{fig3barm1}(b).
There are three two-fold rotation axes and three vertical mirrors, which are related to each other by three-fold rotational symmetry.
We note that the proposed space group $R\bar{3}$ at low temperature \cite{park2016emergence,glamazda2017relation,mu2022role} is expected to make the tensor components more complicated \cite{akgoz1975space,utermohlen2021symmetry} due to the ferroaxial nature of the point group \cite{gopalan2011rotation,hlinka2016symmetry}.
The absence of $\kappa_{xy}(B_y)$ in $\alpha$-\ce{RuCl3} \cite{yokoi2021half}, which is allowed in the $\bar{3}$ system, validates this setting.
According to the symmetry conditions for the in-plane thermal Hall effect, the form of $\hat{\lambda}$ is obtained from Eq. (\ref{rho2D}) by replacing $\rho$ with $\lambda$.

First, we prepare a sample for the measurement with $J_x$ and $B_x$ (Fig. \ref{fig3barm1}(c)).
We obtain the field dependence of $\lambda^{e}_{xx}(B_x)$ ($\propto \Delta T_x$), and $\lambda^{o}_{yx}(B_x)$ ($\propto \Delta T_y$ with the antisymmerization by field), but not $\lambda^{e}_{yy}(B_x)$.
Instead, we measure the anisotropy $\lambda^{e}_{xx}-\lambda^{e}_{yy}$ by rotating the magnetic field \textit{in-situ} around the $z$ axis by $120^{\circ}$ and perform the same measurement (see left bottom of Fig. \ref{fig3barm1}(c)).
This configuration is similar to the planar Hall effect measurement, and the presence of the field-even off-diagonal component can be known, that is in fact proportional to $\lambda^{e}_{xx}-\lambda^{e}_{yy}$.

Due to the three-fold rotational symmetry of the crystal, we can accurately predict the form of $\hat{\lambda}(B_{120^{\circ}})$ using Eq. (\ref{fig3barm1}) with the rotation matrix, $\mathcal{R}_{120^{\circ}}$,
\begin{multline}
\hat{\lambda}_{2D}(B_{120^{\circ}})=\mathcal{R}_{120^{\circ}}^{-1}\hat{\lambda}_{2D}(B_x)\mathcal{R}_{120^{\circ}}\\
=
\begin{pmatrix}
(\lambda^{e}_{xx}+3\lambda^{e}_{yy})/4 & -\lambda^{o}_{yx}+\sqrt{3}(\lambda^{e}_{xx}-\lambda^{e}_{yy})/4\\
\lambda^{o}_{yx}+\sqrt{3}(\lambda^{e}_{xx}-\lambda^{e}_{yy})/4 & (3\lambda^{e}_{xx}+\lambda^{e}_{yy})/4
\end{pmatrix}
\end{multline}
Accordingly, we obtain $\lambda^{o}_{yx}+\sqrt{3}(\lambda^{e}_{xx}-\lambda^{e}_{yy})/4$ as the off-diagonal component.
By measuring the field-dependence, we separate the field-symmetric and antisymmetric components in the $yx$ component, respectively, as 
\begin{equation}
\lambda^{S}_{yx}(B_{120^{\circ}})=\sqrt{3}(\lambda^{e}_{xx}-\lambda^{e}_{yy})/4+\delta
\end{equation}
and
\begin{equation}
\lambda^{AS}_{yx}(B_{120^{\circ}})=\lambda^{o}_{yx}.
\end{equation}
We introduce the $\delta$ term in the first equation, which is proportional to the longitudinal component owing to the possible misalignment of the Hall electrodes.

The $\delta$ term can be eliminated by performing the same measurement with the field rotation by $-120^{\circ}$ (left top of Fig. \ref{fig3barm1}(c)).
The field-symmetric contribution to the off-diagonal thermal Hall resistivity is given as
\begin{equation}
\lambda^{S}_{yx}(B_{-120^{\circ}})=-\sqrt{3}(\lambda^{e}_{xx}-\lambda^{e}_{yy})/4+\delta.
\end{equation}
We note that the intrinsic component ($\propto \lambda^{e}_{xx}-\lambda^{e}_{yy}$) is reversed from that for $B_{120^{\circ}}$, while the extrinsic $\delta$ term is not.
An antisymmetrization with respect to $\pm 120^{\circ}$ can be defined as $\lambda^{S'}_{yx}(|B|) =(\lambda^{e}_{yx}(B_{120^{\circ}})-\lambda^{e}_{yx}(B_{-120^{\circ}}))/2$.
We obtain $\lambda^{S'}_{yx}=\sqrt{3}(\lambda^{e}_{xx}-\lambda^{e}_{yy})/4$, which provides the anisotropy between $\lambda^{e}_{xx}(B_x)$ and $\lambda^{e}_{yy}(B_x)$.
Here, we note that the field dependence of $\lambda^{e}_{xx}$, $\lambda^{e}_{yy}$, and $\lambda^{o}_{yx}$ in each equation are identical to those of $B_x$ due to the three-fold rotational symmetry unless the demagnetization effect is significant.

The measurement can be done within a single-device by using an appropriate sample stage with rotators.
The demonstration of $\lambda^{S'}_{yx}=0$ gives a proof for $\lambda^{e}_{xx}(B_x)=\lambda^{e}_{yy}(B_x)$, which justifies the approximate form of Eq. (\ref{kappaxypseudo}) for estimating the thermal Hall conductivity in $\alpha$-\ce{RuCl3}.
Otherwise, the exact form of Eq. (\ref{kappaxyprecise}) has to be used to accurately estimate $\kappa^{o}_{xy}$ in the spin-disordered regime, suggesting the need for a correction of the reported half-integer value.

The other approach to verify $\lambda^{e}_{xx}=\lambda^{e}_{yy}$ in $\alpha$-\ce{RuCl3} is to measure two different samples (two-divice method), and to utilize the thermal Hall effect itself as a sample quality check.
This can be done by preparing two high-quality samples that show quantized $\kappa^{o, \text{approx}}_{xy}/T=\frac{1}{2}K_0/d$ in individual setups for the heat current parallel to the $x$ and $y$ axis (Figs. \ref{figFive}(b) and \ref{figFive}(c)), respectively.
For a sample in the $J_x$ setup (sample X), we measure
\begin{equation}
\kappa^{\text{X,approx}}_{xy}(B_x)=\lambda^{\text{X}}_{yx}/(\lambda^{\text{X}}_{xx})^2.
\end{equation}
With a sample in the $J_y$ setup (sample Y), we obtain
\begin{equation}
\kappa^{\text{Y,approx}}_{y(-x)}(B_x)=\lambda^{\text{Y}}_{(-x)y}/(\lambda^{\text{Y}}_{yy})^2.
\end{equation}
Here, we omit e and o for simplicity.

If we find that the thermal Hall plateaus are quantized to $\kappa^{\text{X,approx}}_{xy}(B_x,T)d/T\simeq \kappa^{\text{Y,approx}}_{y(-x)}(B_x,T)d/T\simeq \frac{1}{2}K_0$, we can confirm their comparable quality.
And, we obtain an indirect proof of $\lambda^{e}_{yy}(B_x,T)\simeq \lambda^{e}_{xx}(B_x,T)$ because
\begin{equation}
1\simeq \sqrt{\kappa^{\text{X,approx}}_{xy}(B_x,T)/\kappa^{\text{Y,approx}}_{y(-x)}(B_x,T)}=\lambda^{\text{Y}}_{yy}/\lambda^{\text{X}}_{xx}.
\end{equation}
On the other hand, if the reproducibility of a different quantization $\kappa^{\text{Y,approx}}_{y(-x)}(B_x)d/T=rK_0$ ($r \neq \frac{1}{2}$) is established for the $J_y$ setup, it may be an artifact due to the approximate forms of the $\kappa_{xy}$ (Eq. (\ref{kappaxypseudo})), and the precise thermal Hall conductivity could be estimated by a geometric mean
\begin{equation}
\kappa^{\text{exact}}_{xy}d/T\simeq \sqrt{\kappa^{\text{X,approx}}_{xy}\kappa^{\text{Y,approx}}_{xy}}d/T=\sqrt{r/2}K_0\neq \frac{1}{2}K_0.
\end{equation}
This would provide an opportunity to reconsider the true quantization value of the thermal Hall conductivity in $\alpha$-\ce{RuCl3}.

The preceding discussion is about the effect due to the possible anisotropy between $\lambda^{e}_{xx}(B_x)$ and $\lambda^{e}_{yy}(B_x)$.
For the quantitative evaluation of the $\kappa_{xy}$ in $\alpha$-\ce{RuCl3}, two additional effects mentioned above also have to be considered.
For the proposed space group $R\bar{3}$ at low temperature, there are twin operations that change the orientation of the \ce{RuCl6} octahedra and reverse the sign of $\kappa_{xy}(B_x)$: the $C_2$ operation perpendicular to the honeycomb lattice ($C_{2z}$) and the mirror operation $m_{yz}$.
It has been pointed out that the $ABAB$-stacking of the honeycomb-layers in the $ABC$-stacking affects the magnetic transition temperature \cite{kubota2015successive,johnson2015monoclinic,cao2016low,bruin2022origin}, and such crystals can be avoided by the magnetization measurements \cite{kasahara2018majorana,yokoi2021half}.
Since the twin domains due to the $C_{2z}$ and $m_{yz}$ are expected to have the same transition temperature, the twinned crystals can only be distinguished, for example, by checking the intensity profile of the electron/x-ray diffraction pattern (see SI Sec. D). 
As for the out-of-plane thermal transport, it has been reported \cite{hentrich2018unusual} to be comparable to the in-plane transport, while the out-of-plane thermal Hall effect $\lambda^{o}_{yz}$ (and also $\lambda^{e}_{xz}$) has not been reported.
Nevertheless, this effect may be negligible as the thermal Hall angle $\lambda^{o}_{yz}/\lambda^{e}_{zz}$ is expected to be significantly small in magnetic insulators \cite{ideue2017giant}.

\section{Conclusion.}
Unlike the normal Hall effect, the in-plane (thermal) Hall effect is a response where the lack of crystal symmetry is essential.
As a result, several factors, including crystal twinning, out-of-plane transport, and in-plane rotational symmetry, have to be taken into account in order to quantitatively evaluate the Hall conductivity under the in-plane field condition.
To accurately extract the Hall conductivity from the measured resistivity components, a formalism beyond the conventionally-used formula is required because the $xy$-plane symmetry is inevitably broken.

In principle, two independent experimental setups are required to measure at least three independent components in the (thermal) resistivity tensor (two-device method).
Possible sample dependence may prevent an accurate conversion to the conductivity tensors, while the values of the Hall resistivity in individual experiments can be used as a check of the equivalence of the sample quality.
Another improved single-device approach using the rotation of the in-plane field is possibly available when the crystal has three-fold rotational symmetry, which is the archetype of the in-plane Hall system for the theoretical consideration of the quantized anomalous Hall effect \cite{liu2013plane,ren2016quantum,sheng2017monolayer,zhong2017plane,zhang2019plane,liang2018anomalous,li2022chern}, and thermal Hall effect of Kitaev-related honeycomb magnets \cite{utermohlen2021symmetry,chern2021sign,zhang2021topological,koyama2021field}.
The protocols presented in this manuscript would be useful to shed additional light on the experimental evidence for the quantization of $\kappa_{xy}$ in $\alpha$-\ce{RuCl3}, and more generally applicable to the evaluation of the unconventional in-plane Hall conductivity of a broader class of quantum materials.

\begin{acknowledgements}
\section{Acknowledment}
T.K. was financially supported by Ministry of Education Culture Sports Science and Technology (MEXT) Leading Initiative for Excellent Young Researchers (JPMXS0320200135), Japan Society for the Promotion of Science (JSPS) KAKENHI Grant-in-Aid for Young Scientists B (No. 21K13874).
We thank L. Ye, T.-h. Arima, J. G. Checkelsky, and S. Kitou for fruitful discussion and comments on the manuscript.
\end{acknowledgements}

\clearpage
\section{Supplementary Materials}

\renewcommand{\thefigure}{S\arabic{figure}}
\setcounter{figure}{0}
\renewcommand{\thesection}{\Alph{section}}
\setcounter{section}{0}
\renewcommand{\theequation}{S\arabic{equation}}
\setcounter{equation}{0}

\section{A. List of point groups for the non-zero in-plane Hall effect}
In the main text, we show that the in-plane Hall effect is allowed only when some symmetries are absent.
As an example, we prove that the two-fold rotation axis along the magnetic field is forbidden (Condition 1, see Fig. 2(a)).
The proofs for the Conditions 2 to 4 (Fig. 2(b)-(c)) are given as follows.

We can apply the corresponding logic to the system with a mirror symmetry perpendicular to the magnetic field (Condition 2, Fig. 2(b)).
The orientation of $J_x$ is reversed by the $m_{yz}$ while the sample configuration (including the induced para- or diamagnetic moment), $E^{o}_y$, and $B_x$ remain the same.
This relationship gives the equation
\begin{equation}
E^{o}_y=\rho^{o}_{yx}(B_x,0,0)\cdot J_x=\rho^{o}_{yx}(B_x,0,0)\cdot (-J_x),
\end{equation}
proving that $\rho^{o}_{yx}=0$.

As for the two-fold rotational axis along the $z$ axis (Condition 3, see Fig. 2(c)), we can prove the absence of the in-plane Hall effect.
This is the case when the crystal has $C_2$, $C_4$, $S_4$, or $C_6$ rotational symmetry along the $z$ axis in zero field.
We assume that although the magnetic field induces para- or diamagnetic moments in the $xy$ plane, they do not reduce the symmetry of the system lower than $\mathcal{T}C_{2z}$, where $\mathcal{T}$ is the time reversal operation.
The $C_{2z}$ rotation of the whole setup changes the upper panel to the bottom in Fig. 2(c).
We obtain the following equations
\begin{equation}
E^{o}_y=\rho^{o}_{yx}(B_x,B_y,0)\cdot J_x,
\end{equation}
\begin{equation}
-E^{o}_y=\rho^{o}_{yx}(-B_x,0,0)\cdot (-J_x).
\end{equation}
Since $\rho^{o}_{yx}(-B_x,0,0)=-\rho^{o}_{yx}(B_x,0,0)$, we prove that $\rho^{o}_{yx}=0$.

For the mirror symmetry in the $xy$ plane (Condition 4, see Fig. 2(d)), the proof is as follows.
Again, we note that the system becomes symmetric against $\mathcal{T}m_{xy}$ operation under the in-plane magnetic field.
The equation for the upper panel in Fig. 2(d) is $E^{o}_y = \rho^{o}_{yx}(B_x,B_y,0)\cdot J_x$, and the mirror operation to the experimental setup gives $E^{o}_y = \rho^{o}_{yx}(-B_x,-B_y,0)\cdot J_x$.
The last equation gives $-E^{o}_y = \rho^{o}_{yx}(B_x,B_y,0)\cdot J_x$ to prove $\rho^{o}_{yx}=0$.

On the basis of the Conditions 1 to 4, we produce the list of the crystal point groups that allow the in-plane Hall effect.
Table~\ref{tablesymm} summarizes such point groups and allowed $\hat{\rho}$ components and field-configurations required to induce the in-plane Hall.
The leading term with respect to the applied field is also shown.
The absence of the orthorhombic point groups in the table indicates that they do not allow the in-plane Hall effect as long as the applied field is in the high-symmetry directions such as the $xy$ plane.

\begin{table*}[t]
\centering
\caption{\label{tablesymm}
Symmetry conditions for the non-zero in-plane Hall effect.
The superscript o is omitted for simplicity.
Point groups that allow the anomalous Hall effect are marked with an asterisk.
}
\begin{tabular}{l*{3}{c}}
\hline
\hline
Point group  &  Symmetry axes    &    Allowed in-plane Hall effect    &    Leading term ($\propto B^n$)\\
\hline
\multicolumn{4}{c}{Para- or diamagnetic materials}\\
\hline
&& \\
$1$, $\bar{1}$ & arbitrary $xyz$ setting & $\rho_{yx}(B_x,B_y,0)$ & $\propto B$ \\
$2$, $m$, $2/m$ & $C_2\parallel y$ or $m\parallel xz$ & $\rho_{yx}(B_x)$, $\rho_{yz}(B_z)$ & $\propto B$\\
$4$, $\bar{4}$, $4/m$ & $C_4$ or $S_4\parallel z$ &$\rho_{xz}(B_x)$, $\rho_{yz}(B_y)$ &$\propto B$ \\
$3$, $\bar{3}$ & $C_3\parallel z$ & $\rho_{xy}(B_x,B_y,0)$, $\rho_{xz}(B_x)$, $\rho_{yz}(B_y)$ & $\propto B$ for $\rho_{xz/yz}$\\
&&&$\propto B^3$ for $\rho_{xy}$\\
$32$, $3m$, $\bar{3}m$ & $C_3\parallel z$, $C_2\parallel y$ or $m \parallel xz$ & $\rho_{xy}(B_x)$ & $\propto B^3$\\
$6$, $\bar{6}$, $6/m$ & $C_6$ or $S_3\parallel z$ & $\rho_{xz}(B_x)$, $\rho_{yz}(B_y)$ & $\propto B$\\
$23$, $m3$ & $C_3\parallel [111] \parallel z$ & $\rho_{xy}(B_x,B_y,0)$, $\rho_{xz}(B_x)$, $\rho_{yz}(B_y)$ & $\propto B$ for $\rho_{xz/yz}$\\
&&&$\propto B^3$ for $\rho_{xy}$\\
$432$, $\bar{4}3m$, $m\bar{3}m$ & $C_3\parallel [111]\parallel z$, $C_2\parallel [1\bar{1}0] \parallel y$ or $m \perp [1\bar{1}0]$ & $\rho_{xy}(B_x)$ & $\propto B^3$\\
\hline
\multicolumn{4}{c}{Magnetic materials}\\
\hline
&& \\
$1^*$, $\bar{1}^*$, $\bar{1}'$ & arbitrary $xyz$ setting & $\rho_{yx}(B_x,B_y,0)$ & $\propto B$ \\
$2^*$, $m^*$, $2/m^*$ & $C_2 \parallel y$ or $m\parallel xz$ & $\rho_{yx}(B_x)$, $\rho_{yz}(B_z)$ & $\propto B$ \\
$2'^*$, $m'^*$, $2'/m'^*$ & $\mathcal{T}C_2 \parallel y$ or $\mathcal{T}m\parallel xz$ & $\rho_{yx}(B_x)$, $\rho_{yz}(B_z)$ & $\propto B$ \\
$2/m'$, $2'/m$ & $(\mathcal{T})C_2 \parallel y$ or $(\mathcal{T})m\parallel xz$ & $\rho_{yx}(B_x)$, $\rho_{yz}(B_z)$ & $\propto B$ \\
$4^*$, $\bar{4}^*$, $4/m^*$ & $C_4$ or $S_4\parallel z$ & $\rho_{xz}(B_x)$, $\rho_{yz}(B_y)$ &$\propto B$ \\
$4'$, $\bar{4}'$, $4'/m$ & $\mathcal{T}C_4$ or $\mathcal{T}S_4\parallel z$ & $\rho_{xz}(B_x)$, $\rho_{yz}(B_y)$ & $\propto B$ \\
$4/m'$, $\bar{4}'/m'$ & $C_4$ or $\mathcal{T}S_4\parallel z$ & $\rho_{xz}(B_x)$, $\rho_{yz}(B_y)$ & $\propto B$ \\
$3^*$, $\bar{3}^*$, $\bar{3}'$ & $C_3\parallel z$ & $\rho_{xy}(B_x,B_y,0)$, $\rho_{xz}(B_x)$, $\rho_{yz}(B_y)$ & $\propto B$ for $\rho_{xz/yz}$\\
&&&$\propto B^3$ for $\rho_{xy}$\\
$32$, $3m$, $\bar{3}m$ & $C_3\parallel z$, $C_2\parallel y$ or $m \parallel xz$ & $\rho_{xy}(B_x)$ & $\propto B^3$\\
$32'^*$, $3m'^*$, $\bar{3}m'^*$ & $C_3\parallel z$, $\mathcal{T}C_2\parallel y$ or $\mathcal{T}m \parallel xz$ & $\rho_{xy}(B_x)$ & $\propto B^3$\\
$\bar{3}'m'$, $\bar{3}'m$ & $C_3\parallel z$, $\mathcal{T}C_2\parallel y$ or $\mathcal{T}m \parallel xz$ & $\rho_{xy}(B_x)$ & $\propto B^3$\\
$6^*$, $\bar{6}^*$, $6/m^*$ & $C_6$ or $S_3\parallel z$ & $\rho_{xz}(B_x)$, $\rho_{yz}(B_y)$ & $\propto B$\\
$6'$, $\bar{6}'$, $6'/m'$ & $\mathcal{T}C_6$ or $\mathcal{T}S_3\parallel z$ & $\rho_{xz}(B_x)$, $\rho_{yz}(B_y)$ & $\propto B$\\
$6/m'$, $\bar{6}'/m$ & $C_6$ or $\mathcal{T}S_3\parallel z$ & $\rho_{xz}(B_x)$, $\rho_{yz}(B_y)$ & $\propto B$\\
$23$, $m3$, $m'3$ & $C_3\parallel [111] \parallel z$ & $\rho_{xy}(B_x,B_y,0)$, $\rho_{xz}(B_x)$, $\rho_{yz}(B_y)$ & $\propto B$ for $\rho_{xz/yz}$\\
&&&$\propto B^3$ for $\rho_{xy}$\\
$432$, $\bar{4}3m$, $m\bar{3}m$ & $C_3\parallel [111]\parallel z$, $C_2\parallel [1\bar{1}0] \parallel y$ or $m \perp [1\bar{1}0]$ & $\rho_{xy}(B_x)$ & $\propto B^3$\\
$4'32$, $\bar{4}'3m'$, $m\bar{3}m'$ & $C_3\parallel [111]\parallel z$, $C_2\parallel [1\bar{1}0] \parallel y$ or $\mathcal{T}m \perp [1\bar{1}0]$ & $\rho_{xy}(B_x)$ & $\propto B^3$\\
$m'3m'$, $m'3m$ & $C_3\parallel [111] \parallel z$, $C_2\parallel [1\bar{1}0] \parallel y$ or $\mathcal{T}m \perp [1\bar{1}0]$  & $\rho_{xy}(B_x)$ & $\propto B^3$\\
\hline
\hline
\end{tabular}
\end{table*}

In the same table, we show the symmetry conditions for the in-plane Hall effect in the magnetic materials, where the time-reversal symmetry is broken due to internal magnetic order parameters.
The low-field expansion of the Hall coefficient for all magnetic point groups has been given in Ref. \cite{grimmer1993general}, which gives the allowed Hall coefficients up to the $B$-linear term.
We present the extension to the arbitrary field region.
Evidently, we can apply the same discussion to the two cases (systems with the symmetry $C_{2x}$, or $m_{yz}$, Condition 1' and 2', corresponding to Figs. 2(a)-(b)) considered in the main text to derive the $\rho^{o}_{yx}(B_x)$ even for magnetic materials.
We also have to assume that the $B_x$ does not induce the further symmetry breaking that lowers the $C_{2x}$ or $m_{yz}$ symmetry, which would be reasonably satisfied in many materials such as ferromagnets with magnetization $M\parallel x$ and two-sublattice antiferromagnets with sublattice moments $S_{\text{A/B}}\parallel x$.
For the antiferromagnets with $S_{\text{A/B}}\parallel y$ or $z$ or more complicated noncolinear magnets, we have to carefully check if the spin canting due to $B_x$ does not lower the original symmetries.

\begin{figure}[t]
	\includegraphics[width =  \columnwidth]{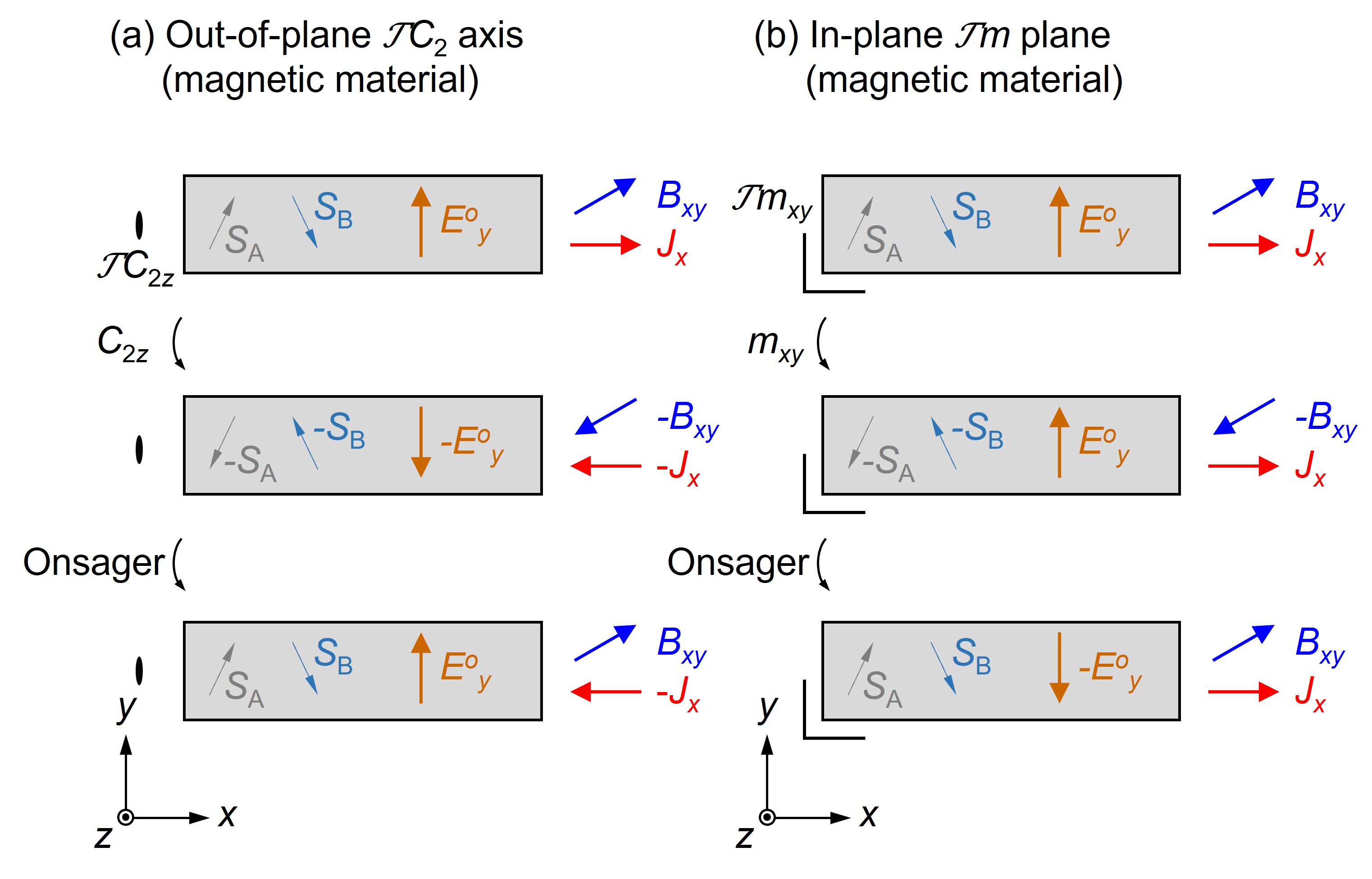}
	\caption{\label{figsymmag}
 Symmetry conditions 3' and 4' for the absence of the in-plane Hall effect in magnetic materials.
 See the caption of Fig. 2 for details.
 The gray and pale blue arrows in the crystal represent the ordered magnetic moments, $S_{\text{A}}$ and $S_{\text{B}}$, which are canted by the applied magnetic field.
 (a) Application of the $C_{2z}$ rotation to the whole setup including the crystal with $\mathcal{T}C_{2z}$ symmetry transforms the top to the middle, which is accompanied with the inversion of $E^o_y$, $B_{xy}$, and $J_x$.
 The Onsager's relation (Eq. (\ref{onsager})) allows the transformation from the middle to the bottom.
 (b) Corresponding figure for the crystal with $\mathcal{T}m_{xy}$ symmetry.
 }
\end{figure}

\begin{figure}[t]
	\includegraphics[width =  \columnwidth]{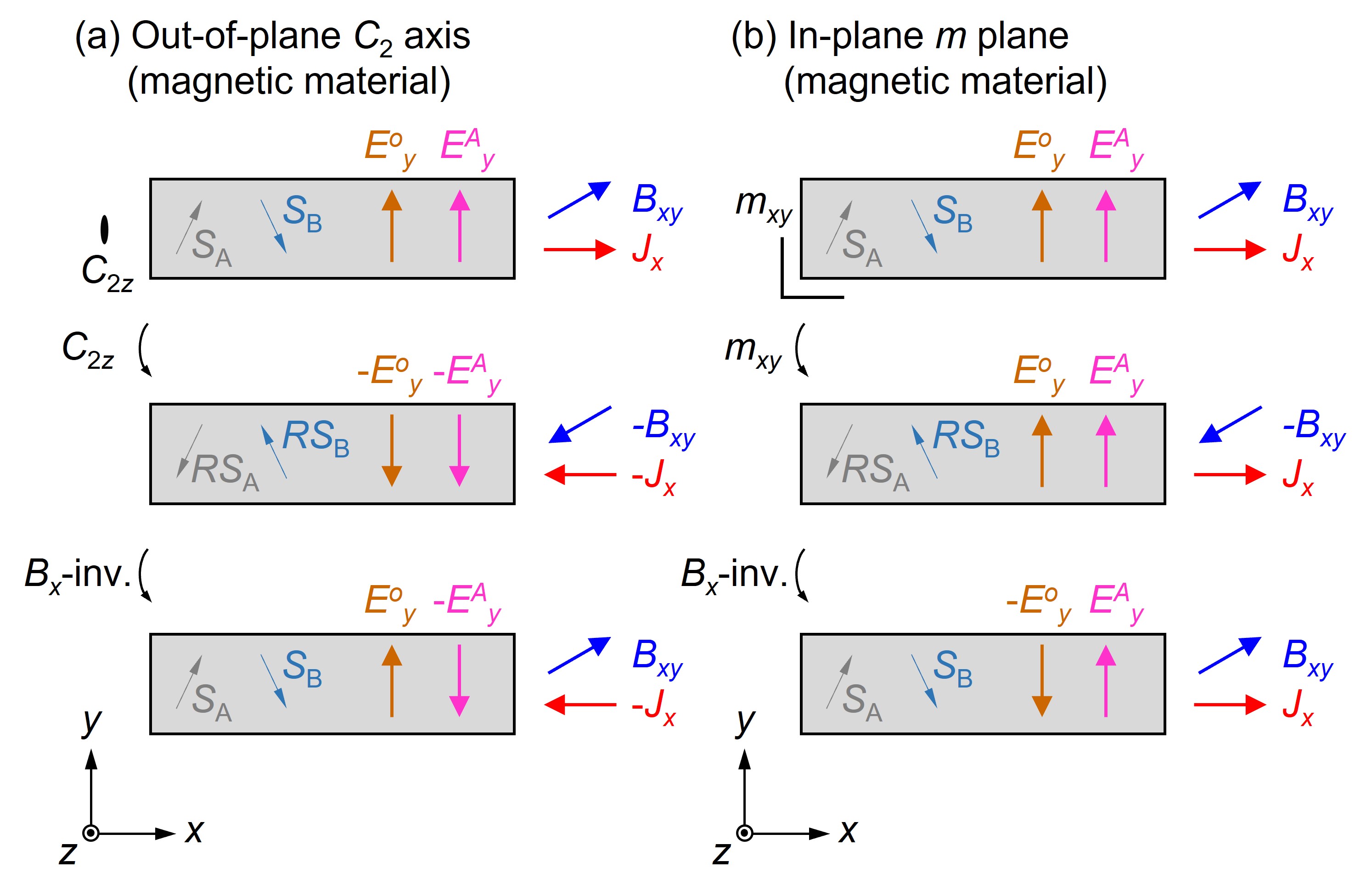}
	\caption{\label{figsymmag2}
 Symmetry conditions 5' and 6' for the absence of the in-plane Hall effect in magnetic materials.
 See the caption of Figs. 2 and \ref{figsymmag} for details.
 The magnetic order parameters, $S_{\text{A}}$ and $S_{\text{B}}$, potentially have the out-of-plane component to induce magnetization along the $z$ axis for the anomalous Hall system.
 The magnetic moments are canted by the applied magnetic field to break the zero-field symmetry.
 For the case of the anomalous Hall system, pink arrows indicate the electric field due to the anomalous Hall effect, $E^{A}_{y}$ under $\pm J_x$.
 We put the $B_{xy}$-field-odd transverse electric field $E^{o}_{y}$ as a tentative in-plane Hall effect, which is proved to be zero.
 (a) The case for the crystal with the $C_{2z}$ rotational symmetry in zero field.
 Application of the $C_{2z}$ rotation to the whole setup transforms the top to the middle, which is accompanied with the inversion of $E^o_y$, $E^{A}_{y}$, $B_{xy}$, and $J_x$.
 The magnetic moments are rotated to $\mathcal{R}S_{\text{A/B}}$.
 The $B_{xy}$-field inversion induces the transformation from the middle to the bottom.
 (b) Corresponding figure for the crystal with the $m_{xy}$ as the zero-field symmetry.
 }
\end{figure}

In addition to the above, there are the other six symmetries need to be considered, i.e., $\mathcal{T}C_2$ along the $z$ axis (Condition 3'), $\mathcal{T}m$ in the $xy$ plane (Condition 4'), $C_{2}$ along the $z$ axis (Condition 5'), $m$ in the $xy$ plane (Condition 6'), $\mathcal{T}C_2$ along the $x$ axis (Condition 7'), and $\mathcal{T}m$ in the $yz$ plane (Condition 8').
We separate them into three parts, and show the pictorial approaches to prove that they give the absence of the in-plane Hall effect.

First, we consider the following two symmetry conditions: $\mathcal{T}C_2$ along the $z$ axis, and $\mathcal{T}m$ in the $xy$ plane on the basis of Figs. \ref{figsymmag}(a)-(b).
\begin{enumerate}
\renewcommand{\labelenumi}{\arabic{enumi}'}
\setcounter{enumi}{2} 
\item $\rho_{yx}^{o}(B_x,B_y,0)$ is zero if there is a $\mathcal{T}C_2$ axis along the $z$ direction under the magnetic field.

We can prove this in three steps.
First, we consider the magnetic material with $\mathcal{T}C_{2z}$ symmetry under a magnetic field along the $x$ axis.
Figure \ref{figsymmag}(a) shows the schematic in-plane field configuration, where we symbolically introduce the internal magnetic moments, $S_{\text{A}}$ and $S_{\text{B}}$, which represent the internal time-reversal symmetry breaking \cite{shtrikman1965remarks}.
In the top panel, we assume that the in-plane Hall electric field $E_y$ is induced by the $J_x$.
We apply the pure two-fold rotation to the whole setup to obtain the middle panel.
We note that the internal magnetic moments are all reversed since $C_{2z}=\mathcal{T}^{-1}$ for the $\mathcal{T}C_{2z}$-symmetric system.
The internal magnetic moments can be reversed by the Onsager's relation (see Ref. [\onlinecite{grimmer1993general}] for detail)
\begin{eqnarray}\label{onsager}
\rho_{yx}^{o}(-S_{\text{A}}, -S_{\text{B}}, -\bm{B}) &=& \rho^{o}_{xy}(S_{\text{A}}, S_{\text{B}}, \bm{B}) \nonumber\\
&=&-\rho^{o}_{yx}(S_{\text{A}}, S_{\text{B}}, \bm{B}).
\end{eqnarray}

The last equation can be expressed by the bottom panel in Fig. \ref{figsymmag}(a).
Comparing the top and the bottom, we find that $E^o_y$ does not change its sign even if we reverse the current direction, which gives the proof for $\rho^{o}_{yx}=0$.

\item $\rho_{yx}^{o}(B_x,B_y,0)$ is zero if there is a $\mathcal{T}m$ in the $xy$ plane under the magnetic field.

We can apply the same discussion for the above.
Application of the pure $m_{xy}$ operation to the whole setup reverses the magnetic field direction in the $xy$ plane as well as the internal magnetic moments (top to middle in Fig. \ref{figsymmag}(b)).
The Onsager's relation gives the transformation from the middle panel to the bottom panel, which gives the opposite sign of $E^o_y$ from the top, i.e., $\rho^{o}_{yx}=0$.
\end{enumerate}

Next, we consider the other two cases: systems with symmetries $C_{2z}$ or $m_{xy}$ in zero field.
In contrast to the nonmagnetic cases (Figs. 2(c)-(d)), the $C_{2z}$ and $m_{xy}$ symmetries are potentially accompanied by the anomalous Hall effect $\rho^{A}_{yx}$ unless additional symmetries forbid the out-of-plane magnetization \cite{grimmer1993general}.
Strictly, the application of the $B_{xy}$ field breaks the symmetry $C_{2z}$ or $m_{xy}$ due to canting of magnetic moments, nevertheless, we can prove that the $B_{xy}$-field induced Hall effect to be zero.
This means that we can even observe the Hall voltage under the in-plane field in such systems, but it cannot be viewed as the intrinsic in-plane Hall effect, but the field-evolution of the anomalous Hall effect $\rho^{A}_{yx}(B_{xy})$.
Sign of $\rho_{yx}(B_{xy})$ only depends on the sign of the order parameters (regarding $S_{\text{A}}$ and $S_{\text{B}}$) responsible for the anomalous Hall effect, i.e., its field dependence is symmetric with respect to the $B_{xy}$ unless the out-of-plane magnetization is flipped.

To prove the above in the pictorial approach, we consider below.
\begin{enumerate}
\renewcommand{\labelenumi}{\arabic{enumi}'}
\setcounter{enumi}{4} 
\item $\rho_{yx}^{o}(B_x,B_y,0)$ is zero if there is a $C_{2}$ along the $z$ axis in zero field unless the applied field flips the spontaneous magnetization.

As shown by the top panel in Fig. \ref{figsymmag2}(a), we consider whether $E^{o}_{y}$ is zero in the magnetic material under $J_x$.
Due to the applied $B_{xy}$, the internal magnetic moments are canted to break the zero-field $C_{2z}$ symmetry.
Here, since the $B$ in the exact $xy$ plane does not lift the degeneracy for the out-of-plane magnetization, we consider the magnetic monodomain state and assume that the possible anomalous Hall electric field $E^{A}_{y}$ is maintained to be the same sign even if the $B_{xy}$ is sweeped from positive to negative.
We apply the $C_{2z}$ rotation to the whole system to obtain the middle panel, where the internal moments are also rotated to $\mathcal{R}S_{\text{A/B}}$ symbolically.
The $B_{xy}$-field inversion transforms the middle to the bottom, where $E^{o}_{y}$ is reversed by definition.
We can safely assume that the internal moments rotate back to $S_{\text{A/B}}$ keeping the sign of $E^{A}_{y}$ as the exact $B_{xy}$ field does not flip the out-of-plane magnetization.
Comparing the top and bottom, we obtain the proof for $\rho^{o}_{yx}=0$.
In contrast to the above rather artificial setup, practically, we might have to foresee that the anomalous Hall effect would be flipped through the sweep $B_{xy}\rightarrow -B_{xy}$ due to potential field-misalignment towards the out-of-plane direction.
This causes the apparent field-odd response of $\rho_{yx}$, but it should not be interpreted as the intrinsic in-plane Hall effect.

\item $\rho_{yx}^{o}(B_x,B_y,0)$ is zero if there is a $m$ in the $xy$ plane in zero field unless the applied field flips the spontaneous magnetization.

We can apply the above discussion as shown in Fig. \ref{figsymmag2}(b).
The mirror inversion with $m_{xy}$ transforms from the top to the middle with the reversal of the internal magnetic moments to $\mathcal{R}S_{\text{A/B}}$ and the $B_{xy}$.
The $B_{xy}$-field inversion gives the bottom panel, where $S_{\text{A/B}}$ goes back to the original configuration.
Only the $E^{o}_{y}$ is remained reversed to give the proof for $\rho^{o}_{yx}=0$.
\end{enumerate}

\begin{figure}[t]
	\includegraphics[width =  \columnwidth]{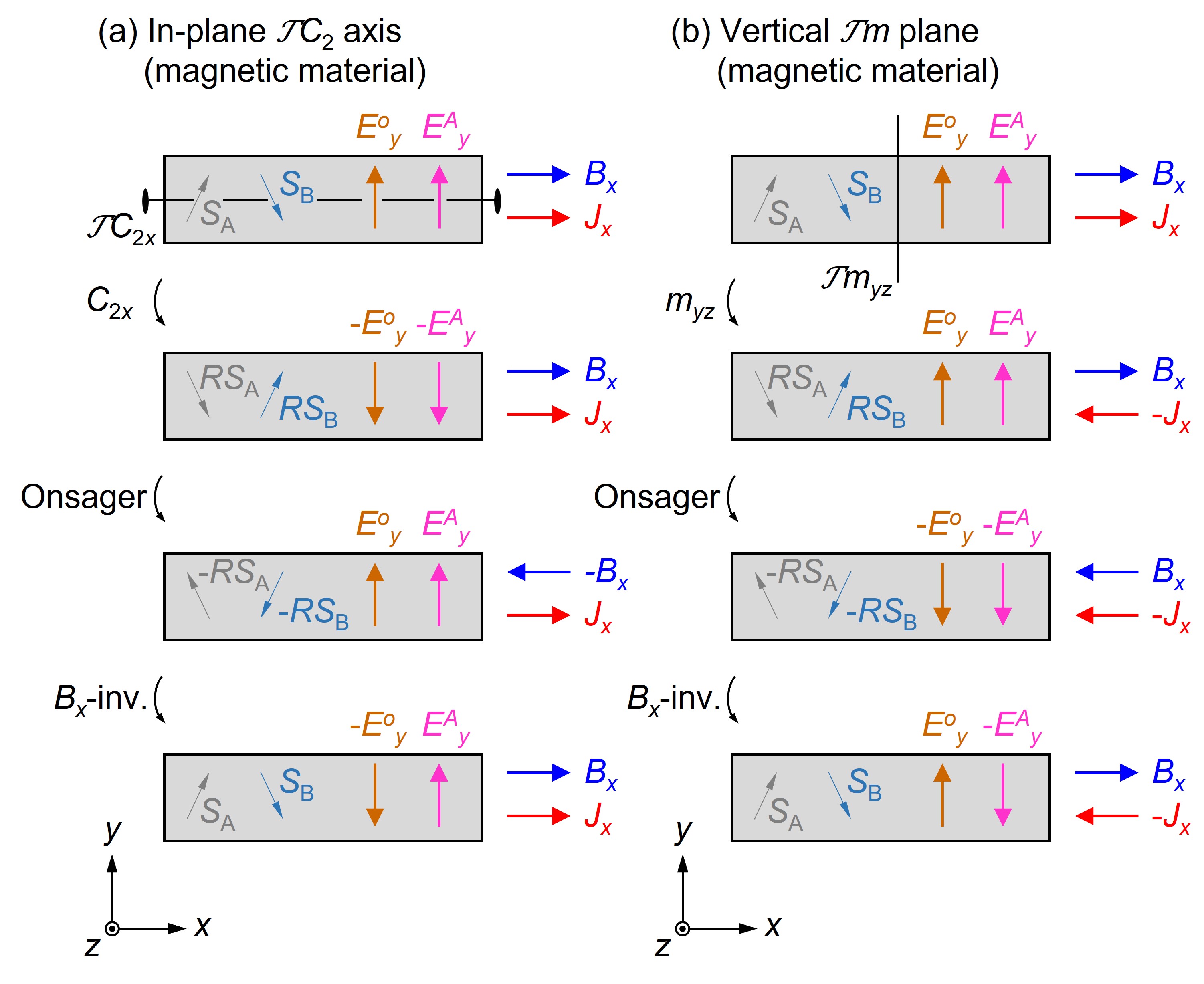}
	\caption{\label{figsymmag3}
 Symmetry conditions 7' and 8' for the absence of the in-plane Hall effect in magnetic materials.
 See the caption of Figs. 2, \ref{figsymmag}, and \ref{figsymmag2} for details.
 (a) The case for the crystal with the $\mathcal{T}C_{2x}$ rotational symmetry in zero field.
 Application of the $C_{2x}$ rotation to the whole setup transforms the top to the second panel, which is accompanied with the inversion of $E^o_y$, and $E^{A}_{y}$.
 The magnetic moments are rotated to $\mathcal{R}S_{\text{A/B}}$.
 The Onsager's relation (Eq. (\ref{onsager})) allows the transformation from the second to the third panel.
 The $B_x$-field inversion induces the transformation from the third panel to the bottom.
 (b) Corresponding figure for the crystal with the $m_{xy}$ as the zero-field symmetry.
 }
\end{figure}

Lastly, we consider the remaining two cases: systems with symmetries $\mathcal{T}C_{2x}$ or $\mathcal{T}m_{yz}$ in zero field.
These two symmetries also potentially allow the spontaneous out-of-plane magnetization and thus the anomalous Hall effect, $\rho^{A}_{yx}$, while the application of the $B_x$ field breaks the symmetries.
We can prove the following two rules in the same manner described above.
\begin{enumerate}
\renewcommand{\labelenumi}{\arabic{enumi}'}
\setcounter{enumi}{6} 
\item $\rho_{yx}^{o}(B_x,0,0)$ is zero if there is a $\mathcal{T}C_{2}$ along the $x$ axis in zero field unless the applied field flips the spontaneous magnetization.

As shown by the top panel in Fig. \ref{figsymmag3}(a), we consider whether $E^{o}_{y}$ is zero in the magnetic material under $J_x$.
Due to the applied $B_x$, the internal magnetic moments are canted to break $\mathcal{T}C_{2x}$ symmetry for the zero field.
Here, since the $B$ along the exact $x$ axis does not lift the degeneracy for the out-of-plane magnetization, we assume the possible anomalous Hall electric field $E^{A}_{y}$ is maintained to be the same sign.
We apply the $C_{2x}$ rotation to the whole system to obtain the second panel, where the internal moments are also rotated to $\mathcal{R}S_{\text{A/B}}$ symbolically.
The application of the Onsager's relation (Eq. (\ref{onsager})) to the second panel leads to the third panel. 
Eventually, the $B_x$-field inversion leads to the bottom panel, where only the $E^{o}_{y}$ is reversed by definition.
We can safely assume that the internal moments rotate back to $S_{\text{A/B}}$ keeping the sign of $E^{A}_{y}$ as the exact $B_x$ field does not flip the out-of-plane magnetization.
This can be know from the fact that, through the transformation from the top panel to the third, the possible out-of-plane magnetization is reversed twice by the $C_{2x}$ and the Onsager's relation to recover into the original configuration.
Comparing the top and the bottom, we obtain the proof for $\rho^{o}_{yx}=0$.

\item $\rho_{yx}^{o}(B_x,0,0)$ is zero if there is a $\mathcal{T}m$ in the $yz$ plane in zero field unless the applied field flips the spontaneous magnetization.

We can apply the same steps as described above to Fig. \ref{figsymmag3}(b).
The top panel is equivalent to the bottom panel, where $E^{o}_{y}$ is unchanged even though the $J_x$ is reversed.
This means the $\rho^{o}_{yx}=0$.
\end{enumerate}

On the basis of the forbidden rules derived above, we can deduce the magnetic point groups that allow the in-plane Hall effect as summarized in Table~\ref{tablesymm} \footnote{In Ref. \cite{tan2021unconventional}, a similar table is also given, where ferromagnetic cases for each point group are considered, while the cubic groups are excluded without sufficient justification.}.
For example, the (magnetic) monoclinic $2/m$ does not show the finite $\rho^{o}_{xz}(B_x,B_y,0)$ aside from the anomalous Hall effect $\rho^{A}_{xz}$ due to $M\parallel y$ because of the presence of the $m_{xz}$ (and $C_{2y}$ as well), while it allows $\rho^{o}_{yx}(B_x)$.
As for the tetragonal point group $4'$ with the $\mathcal{T}C_4$ rotation along the $z$ axis, we can prove $\rho^{o}_{yx}(B_x,B_y,0)=0$ because the operation of the $\mathcal{T}C_4$ rotation twice is equivalent with $C_2$ along the $z$ axis.
We note that $\mathcal{T}C_4$ does not allow $\rho^{A}_{yx}$ either in contrast to the $2/m$ case.
The in-plane Hall effect is allowed for $\rho^{o}_{xz}(B_x)$, but forbidden for $\rho^{o}_{xz}(B_z)$ because $B_z$ is parallel to the $\mathcal{T}C_4$ axis.

Similarly to the crystal twin discussed in the main text, different types of twin is expected in magnetic system, i.e., magnetic domains with respect to the time reversal.
In contrast to the former, the reversal of the magnetic moments does not affect the sign of the in-plane Hall effect.
We put the in-plane Hall resistivity as $\rho^{o}_{yx}(S_{\text{A}}+\delta s_{\text{A}}, S_{\text{B}}+\delta s_{\text{B}},B_x)$, where $S_{\text{A/B}}$ are the symbolic sublattice moments and $\delta s_{\text{A/B}}$ are the $B_x$-induced canting towards the field (see Ref. [\onlinecite{grimmer1993general}] for detail).
We apply the Onsager's relation Eq. (\ref{onsager}) to obtain
\begin{eqnarray}
\lefteqn{\rho^{o}_{yx}(S_{\text{A}}+\delta s_{\text{A}}, S_{\text{B}}+\delta s_{\text{B}},B_x)} \nonumber \\
 & = & \rho^{o}_{xy}(-S_{\text{A}}-\delta s_{\text{A}}, -S_{\text{B}}-\delta s_{\text{B}},-B_x)\nonumber \\
 & = & -\rho^{o}_{yx}(-S_{\text{A}}-\delta s_{\text{A}}, -S_{\text{B}}-\delta s_{\text{B}},-B_x)\nonumber \\
 & = & \rho^{o}_{yx}(-S_{\text{A}}+\delta s_{\text{A}}, -S_{\text{B}}+\delta s_{\text{B}},B_x).
\end{eqnarray}
In the last equation, we assume that the $B_x$-field inversion only cause the sign change of $\delta s_{\text{A/B}}$.
Comparing the first and the last equation, we obtain that the sign of the sublattice magnetic moments does not affect the in-plane Hall effect.

The magnetic half-Heusler compound DyPtBi is an example showing the in-plane Hall effect \cite{chen2022unconventional}.
The crystal structure belongs to $\bar{4}3m$ in a paramagnetic state and the antiferromagnetic order breaks the time reversal symmetry.
The in-plane Hall effect is allowed in the $(111)$ plane with the current along the $[1\bar{1}0]$ axis and the magnetic field is away from the current direction as long as the mirror symmetry in the $(1\bar{1}0)$ plane is maintained.
The observed signal for the field parallel to the current ($J\parallel B\parallel [1\bar{1}0]$) suggests a symmetry breaking of $m\perp [1\bar{1}0]$ due to a possible field-induced canted spin texture.
Further analysis of the magnetic structure would be useful to identify the symmetry breaking that enhances the in-plane Hall effect in this system.

\section{B. Effect of out-of-plane transport to the validity of $\hat{\sigma}_{2D}$ and $\hat{\rho}_{2D}$}
In this section, we discuss how to justify the approximation of the $2\times 2$ matrices for the $3\times 3$ matrices of $\hat{\rho}$ and $\hat{\sigma}$.
We provide the explicit form of the $\hat{\rho}$ in Eq. (5), as the inverse matrix of $\hat{\sigma}$ in Eq. (4):
\begin{widetext}
\begin{equation}
\hat{\rho}_{2/m}(B_{x})=\frac{1}{\Delta_{\sigma}}
\begin{pmatrix}
\sigma^{e}_{yy}\sigma^{e}_{zz}+(\sigma^{o}_{yz})^{2} & -\sigma^{o}_{xy}\sigma^{e}_{zz}-\sigma^{o}_{yz}\sigma^{e}_{xz} & \sigma^{o}_{xy}\sigma^{o}_{yz}-\sigma^{e}_{yy}\sigma^{e}_{xz}\\
\sigma^{o}_{xy}\sigma^{e}_{zz}+\sigma^{o}_{yz}\sigma^{e}_{xz} & \sigma^{e}_{xx}\sigma^{e}_{zz}-(\sigma^{e}_{xz})^{2} & -\sigma^{o}_{yz}\sigma^{e}_{xx}-\sigma^{o}_{xy}\sigma^{e}_{xz}\\
\sigma^{o}_{xy}\sigma^{o}_{yz}-\sigma^{e}_{yy}\sigma^{e}_{xz} & \sigma^{o}_{yz}\sigma^{e}_{xx}+\sigma^{o}_{xy}\sigma^{e}_{xz} & \sigma^{e}_{xx}\sigma^{e}_{yy}+(\sigma^{o}_{xy})^{2}
\end{pmatrix}
\end{equation}
\end{widetext}
where $\Delta_{\sigma}$ is expressed as
\begin{multline}
\Delta_{\sigma} =\sigma^{e}_{xx}\sigma^{e}_{yy}\sigma^{e}_{zz}+\sigma^{o}_{xy}\sigma^{o}_{yz}\sigma^{e}_{xz}+\sigma^{o}_{xy}\sigma^{o}_{yz}\sigma^{e}_{xz}\\
-\sigma^{e}_{yy}(\sigma^{e}_{xz})^{2}+(\sigma^{o}_{yz})^{2}\sigma^{e}_{xx}+(\sigma^{o}_{xy})^{2}\sigma^{e}_{zz}
\end{multline}
Here, we assume that the field-odd Hall components, $\sigma^{o}_{xy}$ and $\sigma^{o}_{yz}$ are small compared to the field-even quantities, and set the quadratic form to be zero, e.g., $(\sigma^{o}_{xy})^2=0$.
The $\Delta_{\sigma}$ is approximated as follows:
\begin{multline}
\Delta_{\sigma} \sim \sigma^{e}_{xx}\sigma^{e}_{yy}\sigma^{e}_{zz}-\sigma^{e}_{yy}(\sigma^{e}_{xz})^{2}\\
=\sigma^{e}_{xx}\sigma^{e}_{yy}\sigma^{e}_{zz}(1-\tan\phi_{ex}\tan\phi_{ez})
\end{multline}
We introduce two pseudo Hall angles, $\phi_{ex}=\arctan (\sigma^{e}_{xz}/\sigma^{e}_{xx})$ and $\phi_{ez}=\arctan (\sigma^{e}_{xz}/\sigma^{e}_{zz})$
The form of $\Delta_{\sigma}$ can be understood as being corrected from the off-diagonal-free form $\sigma^{e}_{xx}\sigma^{e}_{yy}\sigma^{e}_{zz}$ by $1-\tan\phi_{ex}\tan\phi_{ez}$.
The correction is negligible when $\tan\phi_{ex}\ll 1$ in a quasi-two-dimensional system, and Eq. (6) is reduced to $\rho^{o}_{yx}=\sigma^{o}_{xy}/(\sigma^{e}_{xx}\sigma^{e}_{yy})$, ensuring a direct proportionality between the in-plane Hall resistivity and conductivity.

Here, we discuss the form of $\hat{\sigma}$ as the inverse matrix of $\hat{\rho}$.
The $\Delta_{\rho}$ is defined as the determinant of $\hat{\rho}$, and is approximated by
\begin{multline}
\Delta_{\rho} \sim \rho^{e}_{xx}\rho^{e}_{yy}\rho^{e}_{zz}-\rho^{e}_{yy}(\rho^{e}_{zx})^{2}\\
=\rho^{e}_{xx}\rho^{e}_{yy}\rho^{e}_{zz}(1-\tan\theta_{ex}\tan\theta_{ez})
\end{multline}
We introduce $\theta_{ex}=\arctan (\rho^{e}_{zx}/\rho^{e}_{xx})$ and $\theta_{ez}=\arctan (\rho^{e}_{zx}/\rho^{e}_{zz})$.
Similar to $\Delta_{\sigma}$, the $\Delta_{\rho}$ is corrected from $\rho^{e}_{xx}\rho^{e}_{yy}\rho^{e}_{zz}$ by $1-\tan\theta_{ex}\tan\theta_{ez}$.
The conductivity tensor components, $\sigma^{e}_{xx}$, $\sigma^{e}_{yy}$, and $\sigma^{o}_{xy}$ are obtained 
\begin{equation}
\sigma^{e}_{xx} \sim 1/[\rho^{e}_{xx}(1-\tan\theta_{ex}\tan\theta_{ez})],
\end{equation}
\begin{equation}
\sigma^{e}_{yy} \sim 1/\rho^{e}_{yy},
\end{equation}
and
\begin{equation}\label{sigmaxy3D}
\sigma^{o}_{xy} \sim (\rho^{o}_{yx}+\rho^{e}_{zx}\rho^{o}_{zy}/\rho^{e}_{zz})/[\rho^{e}_{xx}\rho^{e}_{yy}(1-\tan\theta_{ex}\tan\theta_{ez})].
\end{equation}
Compared to the quasi-two-dimensional forms (Eqs. (10) and (11)), both $\sigma^{e}_{xx}$ and $\sigma^{o}_{xy}$ gets a correction in the denominator due to the crystalline planar Hall effect ($\rho^{e}_{zx}$), which could be negligible in quasi-two-dimensional systems for $\tan \theta_{ez}\ll 1$, i.e., $\rho^{e}_{zz}$ is large.
As for the $\sigma^{o}_{xy}$, the numerator in Eq. (\ref{sigmaxy3D}) also gets a correction by the leakage from the Hall effect $\rho^{o}_{zy}$ in the $yz$ plane.
The two-dimensional approximation of the $\sigma^{o}_{xy}$ (Eq. (11)) is valid only if $\rho^{e}_{zx}\rho^{o}_{zy}/\rho^{e}_{zz}$ is negligible.
This is realized in quasi-two-dimensional systems with low out-of-plane carrier (electrons/phonons/magnons, etc.) mobility giving a small out-of-plane normal-Hall-angle $\rho^{o}_{zy}/\rho^{e}_{zz}$.

\section{C. Symmetry conditions for non-zero $\rho^{e}_{zx}$}
\begin{figure}[t]
	\includegraphics[width =  \columnwidth]{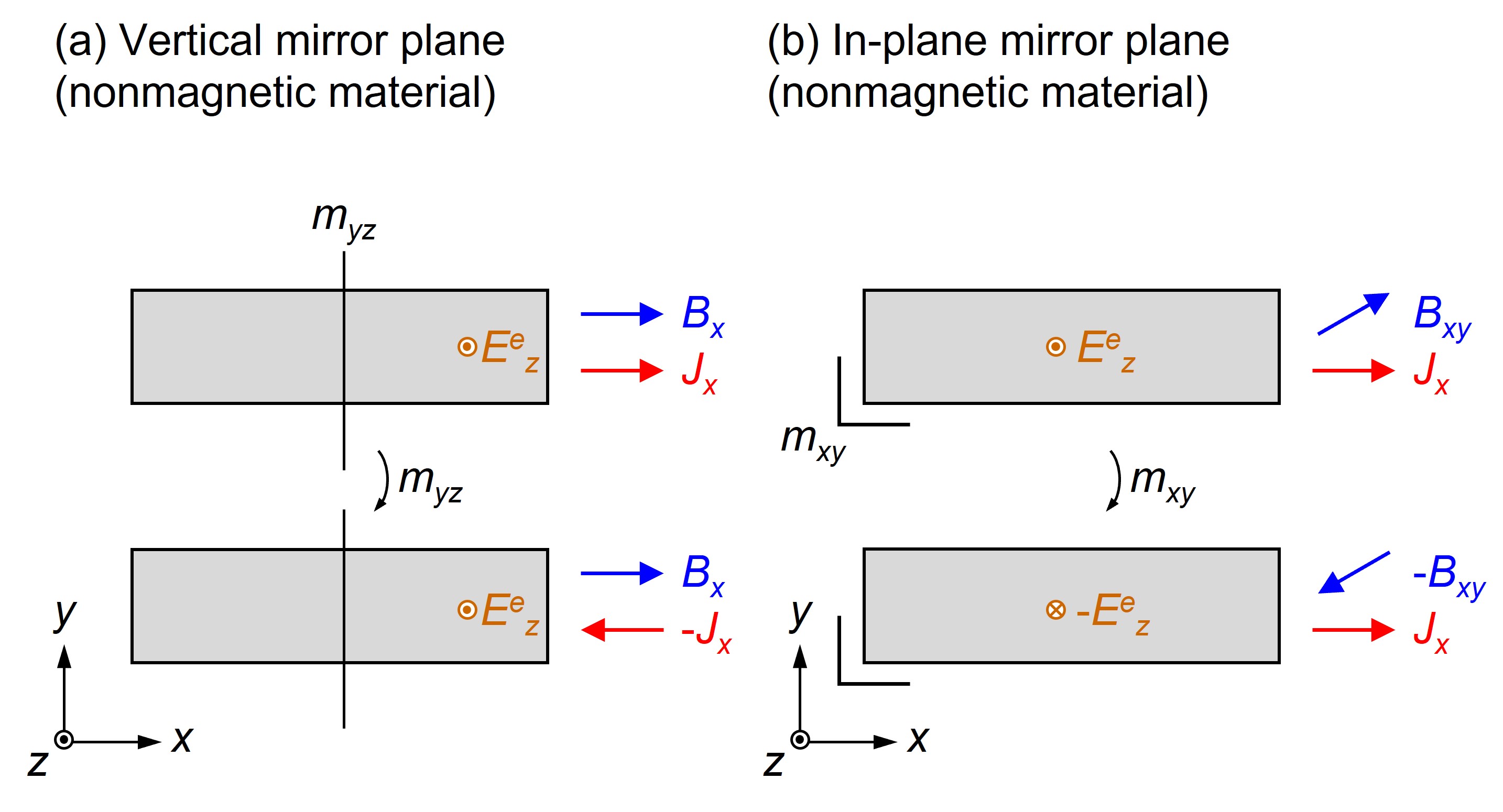}
	\caption{\label{figcph}
 Symmetry conditions for the absence of the crystalline planar Hall effect.
 See the caption of Fig. 2 for details.
 The orange circle with a dot or a cross represents a tentative crystalline planar Hall electric field $\pm E^e_z$ proportional to $\rho^{e}_{zx}$, which is proved to be zero.
 (a) The mirror inversion $m_{yz}$ transforms the top to the bottom, which is accompanied with the inversion of $J_x$, but $B_x$ and $E^e_z$ remain intact.
 (b) Corresponding figure for the crystal with the $m_{xy}$ symmetry.
 }
\end{figure}

As suggested in the main text, the out-of-plane transport inevitably contaminates the in-plane Hall signal through the potentially finite $\sigma^{e}_{xz}$.
Henceforth, we refer to $\sigma^{e}_{xz}$ ($\rho^{e}_{zx}$) as the crystalline planar Hall conductivity (resistivity). 
Indeed, we can prove that the conditions for the non-zero $\rho^{e}_{zx}(B_x,B_y,0)$ (and $\sigma^{e}_{xz}(B_x,B_y,0)$) are equivalent to those for the $\rho^{o}_{yx}(B_x,B_y,0)$ by using the pictorial approaches regarding Figs. 2 and \ref{figsymmag}-\ref{figsymmag3}.

To avoid repeating similar explanations, we show only representative cases.
One example is a vertical mirror symmetry in the $yz$ plane (see Fig. \ref{figcph}(a)).
Here, we consider that a tentative out-of-plane electric field $E^e_z$ is induced by the application of $J_x$, which is proved to be zero.
The mirror operation transforms the system from the top to the bottom, where the $B_x$ and $E^e_z$ remain intact, but the $J_x$ is reversed.
This relationship is expressed as $E^e_z=\rho^{e}_{xz}(B_x,0,0)\cdot J_x =\rho^{e}_{xz}(B_x,0,0)\cdot (-J_x) $, and thus $\rho^{e}_{xz}(B_x,0,0)=0$.
The other example is a horizontal mirror symmetry in the $xy$ plane (see Fig. \ref{figcph}(b)).
The mirror inversion reverses $E^e_z$, and $B_{xy}$ to give the equation $E^e_z=\rho^{e}_{xz}(B_x,B_y,0)\cdot J_x =-\rho^{e}_{xz}(-B_x,-B_y,0)\cdot J_x$.
We note that $\rho^{e}_{xz}(B_x,B_y,0)=\rho^{e}_{xz}(-B_x,-B_y,0)$, and thus obtain the proof of $\rho^{e}_{xz}(B_x,B_y,0)$.
We can straightforwardly prove the other conditions for the absence of $\rho^{e}_{xz}(B_x,B_y,0)$.

The leading order in terms of $B$ for $\rho^{e}_{xz}$ (and $\sigma^{e}_{xz}$ as well) depends on the crystal symmetry.
Triclinic and monolcinic systems are $\rho^{e}_{xz}\neq 0$ even in zero field.
Other point groups start from $\rho^{e}_{xz} \propto B^2$ since it originates from the anisotropy of magnetoresistance, which is zero at $B=0$.

\section{D. Thermal transport measurements and twin operations of $\alpha$-\ce{RuCl3}}
In Table~\ref{tablerucl3}, we show a list of published literature reporting the thermal transport properties of $\alpha$-\ce{RuCl3}.
Each raw summarizes the experimental conditions (heat current and magnetic field direction), the equations, either Eq. (13) or (14), used to estimate the thermal Hall conductivity, and growth methods for single crystals.

\begin{figure}[t]
	\includegraphics[width =  \columnwidth]{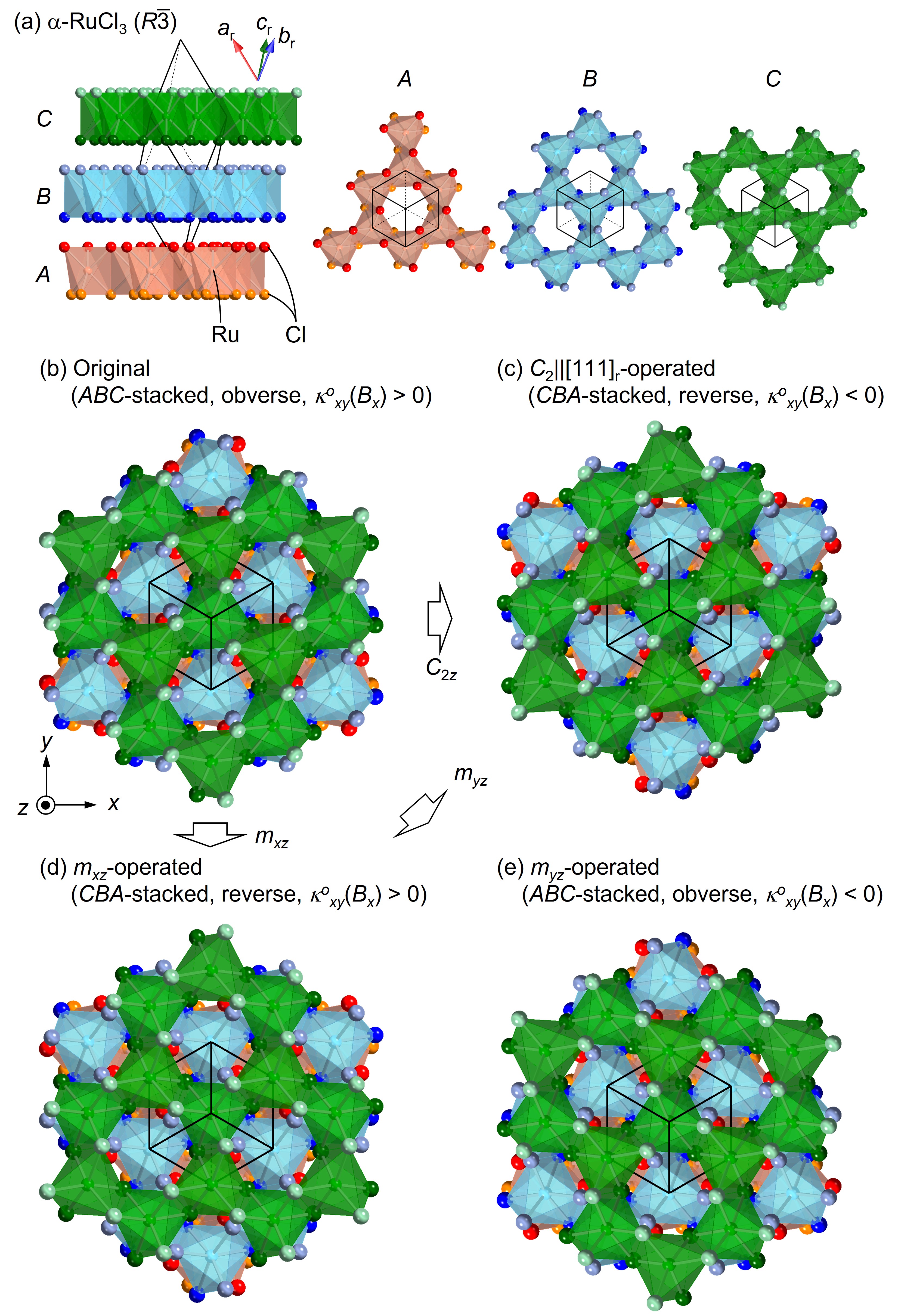}
	\caption{\label{figrucl3twin}
(a) Left: The proposed crystal structure of $\alpha$-\ce{RuCl3} at low temperature \cite{park2016emergence}, which belongs to the $R\bar{3}$ space group.
Right: top view of the three types of Cl-Ru-Cl honeycomb layer of different orientations denoted as $A$, $B$, and $C$.
Displacement of the atomic position of chlorine from the symmetric position is exaggerated for visibility.
The black line is the unit cell in the rhombohedral axes.
(b) Top view of the original $R\bar{3}$ crystal structure, characterized by the $ABC$ stacking and the obverse setting.
The Cartesian coordinate, $xyz$, is defined, and the expected thermal Hall conductivity $\kappa^{o}_{xy}(B_x)$ is set to be positive.
(c)-(d) Possible twin domains produced by the twin operations: $C_{2z}$ for (c), $m_{xz}$ for (d), and $m_{yz}$ for (e).
The stacking order ($ABC$ or $CBA$), the lattice setting (obverse and reverse), and the sign of the $\kappa^{o}_{xy}(B_x)$ are also shown.
 }
\end{figure}

Figure \ref{figrucl3twin}(a) shows the proposed crystal structure of the $R\bar{3}$ \ce{RuCl3} \cite{park2016emergence,glamazda2017relation,mu2022role}, which is described by the stacking of Cl-Ru-Cl layers along the $[111]_{\text{r}}$ (r denotes the rhombohedral lattice) in the order of $A$, $B$, and $C$ (see Fig. \ref{figrucl3twin}(b)).
Each layer can be completely superimposed by a pure spatial translation of $[100]_{\text{r}}$.
As the $R\bar{3}$ lattice does not have the symmetries, $C_{2z}$, $m_{xz}$, and $m_{yz}$, three different types of twin domains associated with the stacking faults can be considered.
Figure \ref{figrucl3twin}(c) shows the crystal structure after the $C_{2z}$ operation.
The $ABC$ stacking is converted to the $CBA$ stacking and the rhombohedral unit cell is changed to the reverse setting (see the orientation of the unit cell).
As we discussed the symmetry condition for the in-plane Hall effect in Fig. \ref{figsymm}(c), the $C_{2z}$ rotation reverses the sign of the $\kappa^{o}_{xy}(B_x)$ $(< 0)$.
A similar discussion can also be applied to the $m_{xz}$ and $m_{yz}$ operations as summarized in Figs. \ref{figrucl3twin}(b) and \ref{figrucl3twin}(c), respectively.
We note that, in contrast to the case of the triangular lattice considered in Fig. \ref{figtwin}(c), the obverse-reverse twinning (twinning by reticular merohedry) is irrelevant for the sign reversal of the $\kappa^{o}_{xy}$.
This occurs between the ferroaxial domains \cite{gopalan2011rotation,hlinka2016symmetry} in the same obverse-reverse settings, which can be identified by electrogyration \cite{aizu1964reversal,gupta2014reinvestigation}, and electron/x-ray diffraction \cite{nord1989order}.

\begin{table*}[t]
\centering
\caption{\label{tablerucl3}
Experimental conditions of the thermal transport measurements for $\alpha$-\ce{RuCl3}.
$\perp z$ for the heat current direction indicates no specification of the crystallographic axis.
Accordingly, the measured thermal conductivity is denoted by $\kappa_{\perp}$ instead of $\kappa_{xx}$ or $\kappa_{yy}$.
There are two ways, exact and approximate, to experimentally estimate $\kappa_{xy}$ according to Eqs. (13) and (14), respectively.
Due to the lack of consensus on the low-temperature crystal structure, we provisionally hypothesize the in-plane isotropy for the $B\parallel z$ measurements (marked with an asterisk).
}
\begin{tabular}{l*{7}{c}}
\hline
\hline
No. & Heat current & Magnetic field &  Measured components  &  Form of $\kappa_{xy}$ & Growth method & Ref.  \\
\hline
1. & $\perp z$ & NA & $\kappa_{\perp}$ & NA & Bridgman & \onlinecite{hirobe2017magnetic} \\
2. & $\perp z$ & $\perp z$, $\parallel J$ & $\kappa_{\perp}$ & NA & vacuum sublimation & \onlinecite{leahy2017anomalous} \\
 & $\perp z$ & $\perp z$, $\perp J$ & $\kappa_{\perp}$ & NA & &  \\
3. & $\perp z$ & $\parallel z$ & $\kappa_{\perp}$, $\kappa_{xy}$ & exact* & Bridgman & \onlinecite{kasahara2018unusual} \\
4. & $\perp z$ & $\perp z$, $\perp J$ & $\kappa_{\perp}$ & NA & CVT & \onlinecite{yu2018ultralow} \\
5. & $\perp z$ & $\perp z$, $\perp J$ & $\kappa_{\perp}$ & NA & CVT/vacuum sublimation & \onlinecite{hentrich2018unusual} \\
 & $\parallel z$ & $\perp z$, $\perp J$ & $\kappa_{zz}$ & NA & & \\
6. & $\parallel x$ & toward $x$ from $z$ by $60^{\circ}$ and $45^{\circ}$ & $\kappa_{xx}$, $\kappa_{xy}$ & approximate & Bridgman & \onlinecite{kasahara2018majorana} \\
7. & $\perp z$ & $\parallel z$ & $\kappa_{\perp}$, $\kappa_{xy}$ & exact* & CVT & \onlinecite{hentrich2019large} \\
8. & $\perp z$ & $\perp z$, $\perp J$ & $\kappa_{\perp}$ & NA & CVT & \onlinecite{hentrich2020high} \\
9. & $\parallel x$ & toward $x$ from $z$ by $45^{\circ}$ & $\kappa_{xx}$, $\kappa_{xy}$ & approximate & Bridgman & \onlinecite{yamashita2020sample} \\
10. & $\perp z$ & $\parallel x$ & $\kappa_{\perp}$, $\kappa_{xy}$ & approximate & CVT & \onlinecite{czajka2021oscillations} \\
 & $\perp z$ & $\parallel y$ & $\kappa_{\perp}$, $\kappa_{xy}$ & approximate & &  \\
11. & $\parallel x$ & $\parallel x$ & $\kappa_{xx}$, $\kappa_{xy}$ & approximate & Bridgman & \onlinecite{yokoi2021half} \\
 & $\parallel x$ & $\parallel y$ & $\kappa_{xx}$, $\kappa_{xy}$ & approximate & & \\
12. & $\parallel x$ & $\parallel z$ & $\kappa_{xx}$, $\kappa_{xy}$ & exact* & CVT & \onlinecite{lefranccois2022evidence} \\
 & $\parallel x$ & toward $x$ from $z$ & $\kappa_{xx}$, $\kappa_{xy}$ & approximate & & \\
13. & $\parallel x$ & $\parallel x$ & $\kappa_{xx}$, $\kappa_{xy}$ & approximate & CVT & \onlinecite{czajka2022planar} \\
14. & $\parallel x$ & NA & $\kappa_{\perp}$ & NA & Bridgman & \onlinecite{bruin2022robustness} \\
15. & $\parallel x$ & $\parallel x$ & $\kappa_{xx}$, $\kappa_{xy}$ & approximate & Bridgman & \onlinecite{suetsugu2022evidence} \\
& $\parallel x$ & $\parallel y$ & $\kappa_{xx}$, $\kappa_{xy}$ & approximate &  &  \\
16. & $\parallel x$ & $\parallel x$ & $\kappa_{xx}$ & NA & Bridgman & \onlinecite{bruin2022origin} \\
 & $\parallel x$ & $\parallel y$ & $\kappa_{xx}$ & NA & Bridgman & \\
 & $\parallel x$ & $\parallel x$ & $\kappa_{xx}$ & NA & CVT &  \\
 & $\parallel x$ & $\parallel y$ & $\kappa_{xx}$ & NA & CVT &  \\
17. & $\parallel x$ & toward $x$ from $z$ & $\kappa_{xx}$, $\kappa_{xy}$ & approximate & Bridgman & \onlinecite{kasahara2022quantized} \\
18. & $\parallel x$ & $\parallel x$ and $\parallel y$ & $\kappa_{xx}$& NA & CVT & \onlinecite{lefranccois2023oscillations} \\
 & $\parallel y$ & $\parallel x$ and $\parallel y$ & $\kappa_{yy}$ & NA && \\
19. & $\parallel x$ & $\parallel x$ & $\kappa_{xx}$, $\kappa_{xy}$ & approximate & CVT & \onlinecite{zhang2023sample} \\
20. & $\parallel x$ & $\parallel x$ & $\kappa_{xx}$, $\kappa_{xy}$ & approximate & CVT & \onlinecite{zhang2023stacking} \\
\hline
\hline
\end{tabular}
\end{table*}

\clearpage
\bibliography{reference}

\end{document}